\documentclass[10pt,twocolumn,aps,prd,amsmath,amssymb,nofootinbib]{revtex4-2}

\usepackage{hyperref}
\usepackage{url}
\usepackage{cleveref}
\usepackage{graphicx}
\usepackage{dcolumn}
\usepackage{bm}
\usepackage[utf8]{inputenc}

\newcommand{\mpl}{M_{\text{pl}}}

\newcommand{\bfE}{{\bf{E}}}
\newcommand{\bfB}{{\bf{B}}}

\newcommand{\bfx}{{\bf{x}}}
\newcommand{\bfk}{{\bf{k}}}

\newcommand{\bfxi}{{\boldsymbol{\xi}}}

\newcommand{\bftheta}{{\boldsymbol{\theta}}}
\newcommand{\bfalpha}{{\boldsymbol{\alpha}}}
\newcommand{\gagg}{g_{a\gamma \gamma}}

\begin{document}

\preprint{APS/123-QED}

\title{Optical Lensing by Axion Stars: \\Observational Prospects with Radio Astrometry}

\author{Anirudh Prabhu}
\email{aniprabhu@stanford.edu}
 \affiliation{Stanford Institute for Theoretical Physics, Stanford University, Stanford, California 94305, USA}

\date{\today}

\begin{abstract}
Axions and axion-like particles (ALPs) are some of the best-motivated dark matter (DM) candidates. Under certain circumstances, large axion fluctuations in the early universe can collapse to form dense configurations called \emph{axion clumps}. The densest axion clumps are metastable states known as \emph{oscillons}. In this paper we propose a new class of observables that exploit the axion's coupling to photons. As a result of this coupling an axion clump acts like an inhomogeneous refractive optical medium--- a lens--- that causes anomalous dispersion of incident electromagnetic waves. The dispersion of electromagnetic waves by axion clumps clearly distinguishes this lensing effect from gravitational lensing. Axion clumps passing in front of background radio sources act as lenses and lead to apparent positional shifts that can potentially be discovered by high-precision radio astrometry missions with the forthcoming Square Kilometer Array (SKA). We discuss the sensitivity of SKA to lensing effects in a variety of axion halo models. While gravitational microlensing surveys have placed strong constraints on the amount of DM that exists in the form of non-luminous astrophysical objects they have been unable to do so for objects in the mass range $[10^{-14}, 10^{-11}] M_\odot$. We find that, over a wide range of parameter space, SKA will be sensitive to optical lensing by oscillons in the mass range that is currently unconstrained by microlensing surveys.  
\end{abstract}

\maketitle


\section{\label{sec:intro}Introduction}

There is overwhelming evidence that a majority of the matter density of the universe exists in a dark sector that interacts very weakly with the Standard Model (SM). The existence of such dark matter (DM) has been supported by observations including the orbital motion of stars and gas clouds in disc galaxies, temperature anisotropies in the cosmic microwave background (CMB), strong gravitational lensing by galaxy clusters, and structure formation. All of the observational evidence to date for the existence of DM relies solely on its gravitational interactions with itself and ordinary matter. Determining the non-gravitational interactions of DM is one of the most important outstanding problems in physics. A general class of DM candidates includes non-luminous, compact massive objects. Strong exclusions have been placed by MACHO \cite{MACHO2001}, EROS \cite{EROSMACHO2007}, OGLE \cite{OGLE2011}, and HSC/Subaru \cite{HSCSubaru2017}, on objects with mass exceeding $10^{-11} \ M_\odot$. For lower masses, constraints exist assuming the compact objects are primordial black holes (PBHs) \cite{PBHEvaporation, PBHEvaporation, PBHNeutronStarCapture, PBHNS2013, PBHNS20132, PBHWD2015, PBHVoyager2018, PBHWill}. While in the low-mass regime, the PBH scenario is strongly constrained, bounds on general compact massive objects are scarce. Femtolensing of gamma ray bursts provides the possibility to probe compact objects with mass less than $10^{-14} M_\odot$, assuming these objects make up more than 10\% of DM \cite{Katz_2018}. As in the case of PBHs, exploiting the microphysical properties of a given compact object may lead to stronger constraints than those from gravitational lensing. 

In this paper, we focus on the possibility that some DM could be in the form of compact massive objects composed of axions. The QCD axion, which arises as a pseudo Nambu-Goldstone boson of a global $U(1)$ Peccei-Quinn (PQ) symmetry spontaneously broken at high energy scale $f_a$, was originally introduced as part of a dynamical solution to the strong CP problem \cite{PQ1,PQ2,WeinbergAxion,WilczekAxion}. The QCD axion can also be produced, via the misalignment mechanism, in the correct abundance to explain DM \cite{Abbott1982, Fischler1982,PRESKILL1983127}. Its prospect as a DM candidate is further strengthened by its feeble interactions with the Standard Model. Many theories beyond the Standard Model, most prominently string theory, predict a plenitude of axion-like particles (ALPs) over a wide range of masses and couplings \cite{Axiverse2010}. These particles can similarly provide good DM candidates. Under certain circumstances, axions can form dense clumps. These clumps form due to the gravitational collapse of large density fluctuations of the axion field. Large fluctuations can be formed if PQ symmetry is broken after inflation, leading to ultradense \emph{axion miniclusters}, or through the \emph{large-misalignment mechanism} \cite{marios2019}. Through mergers and/or accretion, these objects can form even denser objects called \emph{solitons} and \emph{oscillons}. The possibility that an $\mathcal{O}(1)$ fraction of dark matter is bound into such structure can have serious implications for direct detection experiments. Associated with the axion clump scenario is a rich set of astrophysical signatures. Among these signatures are: bright radio flashes sourced by axion-photon conversion in axion star-neutron star collisions \cite{iwazaki2014, Iwazaki2015, Raby2016, bai2018, prabhu2020resonant}\footnote{Some proposals suggest that these are responsible for Fast Radio Bursts.} and gravitational microlensing, picolensing, and femtolensing \cite{fairbairn2018microlensing, KolbTkachevPicoFemtolensing}. Another fascinating possibility is that, in certain regions of parameter space, axion clumps may frequently collide with the Earth giving large transient boosts in the local DM density, which could potentially be observed in existing direct detection experiments \cite{marios2019}. Recently, data from searches for compact massive objects has been used to place constraints on the fractional abundance of axion miniclusters \cite{Fairbairn2017, fairbairn2018microlensing}. 

In this article we propose a new class of observables that exploit unique features of the axion-photon coupling. Due to this coupling, an axion clump acts as an inhomogeneous refractive medium, which can lead to effects such as time delays, spectral distortions, and refraction \cite{mcdonald2019optical, McDonald2020}. These effects are dispersive in nature, allowing them to be distinguished from purely gravitational effects. The main result of this paper is that the refraction of light by inhomogeneous axion configurations due to axion-photon interactions can have unique observable lensing signatures. We refer to such lensing as \emph{optical lensing by axion stars} (OLAS)\footnote{We call this effect \emph{optical lensing} to specify that the lensing object---the axion star--- is an \emph{optical medium} that allows electromagnetic waves across the entire spectrum to pass through it. In this paper we are interested in optical lensing primarily in the radio band of the electromagnetic spectrum. }, to distinguish it from canonical gravitational lensing. One consequence of OLAS is that it can lead to frequency-dependent shifts in the apparent positions of background radio sources, that can potentially be observed by astrometric radio missions such as the Square Kilometer Array (SKA). 

We outline a program to detect such positional shifts using SKA and evaluate its sensitivity to various axion clump models. We find that OLAS is observable for only axion configurations with very high central amplitude--- solitons and oscillons. Over a wide range of parameter space, we find that the OLAS program is sensitive to oscillon masses covering the unconstrained mass range $M_\text{osc} \in \left[10^{-14} M_\odot, 10^{-11} M_\odot \right]$, even for very small fractional abundances of oscillons. This could provide one of the first constraints on compact massive objects in the above mass range.

The paper is organized as follows. In Sec. \ref{sec:AEM}, we discuss the general properties of the axion-photon interaction and derive the effect of light deflection by axion clumps. In Sec. \ref{sec:clumpmodels}, we review some well-motivated models of axion clumps and discuss their formation mechanisms and viability as dark matter candidates. In Sec. \ref{sec:signatures} we outline some of the observable signatures of axion clumps and present the sensitivity of surveys like the Square-Kilometer Array (SKA) for detection. In Sec. \ref{sec:conclusions} we provide some concluding remarks. 

\section{Notation and Conventions \label{sec:convention}}

Unless otherwise stated, we work in natural units $\hbar = c = 1$ and use the Planck mass $\mpl = \sqrt{\frac{\hbar c}{ G_N}}$, where $G_N$ is the gravitational constant. We assume an average dark matter density of $\rho_{_{\text{DM}, 0}} = 1.1 \times 10^{-6}$ GeV/cm$^3$ and a local dark matter density of $\rho_{_{\text{DM},\odot}} = 0.4$ GeV/cm$^3$. We also adopt the $(-+++)$ metric convention. 

\section{Light Refraction in an Axion Background \label{sec:AEM}} 

Interactions between the axion field, $a(x)$, and the Maxwell field are described by the Lagrangian

\begin{align}
    \mathcal{L} \supset -{\gagg \over 4} a \tilde{F}^{
    \mu \nu} F_{\mu \nu}  \label{eqn:action}
\end{align}

\noindent where $F_{\mu \nu} = \partial_\mu A_{\nu} - \partial_\nu A_{\mu}$, $\tilde{F}^{\mu \nu} =\varepsilon^{\mu \nu \alpha \beta} F_{\alpha \beta}/2 $. The axion photon coupling $\gagg$ is usually related to $f_a$ by $\gagg = C_\gamma \alpha_{_\text{EM}}/f_a$ where $\alpha_{_\text{EM}}$ is the electromagnetic fine-structure constant and $C_\gamma$ is an $\mathcal{O}(1)$ constant, although certain models allow for enhanced axion-photon couplings \cite{photophilic2016}, in which case $C_\gamma$ could be arbitrarily large. Maxwell's equation in an axion background are modified as follows.


\begin{align}
\quad \partial_\mu F^{\mu \nu} + (\partial_\mu \theta) \tilde{F}^{\mu \nu} = 0, \quad F_{[\mu \nu, \rho]} = 0 \label{eqn:maxwell}
\end{align}

\noindent where we have defined $\theta(x) \equiv \gagg a(x)$ for notational convenience and we have used the fact that the second equation in (\ref{eqn:maxwell}) is equivalent to $\partial_\mu \tilde{F}^{\mu \nu} = 0$. Approximate plane wave solutions exist in the limit that the axion field is slowly varying in space and time compared to the photon field. Formally, this WKB approximation amounts to the following simplification: $\partial_\mu \partial_\nu \theta/ \partial \theta \ll \partial_\mu F / F$. This motivates the ansatz $F_{\mu \nu} = \mathcal{R}\left\{ A_{\mu \nu} e^{i S} \right\}$ where $A_{\mu \nu}$ is the slowly varying amplitude and $S$ the rapidly varying phase of $F_{\mu \nu}$. In analogy with the plane wave, we may define a frequency and wavenumber 

\begin{align*}
    \omega \equiv - {\partial S \over \partial t}, \quad \bfk \equiv \nabla S 
\end{align*}

Using this ansatz and working with the electric and magnetic fields, (\ref{eqn:maxwell}) takes the form ${\bf{D}}(\omega, \bfk) (\bfE, \bfB)^T = 0$. The dispersion relation is determined by setting det$({\bf{D}}) = 0$. The eigenvalues of ${\bf{D}}$ are given by

\begin{align}
    &D^{\pm} = -\omega^2 + \bfk^2 \pm \nonumber \\
    &\left[ \left(\omega \dot{\theta} + \bfk \cdot \nabla \theta \right)^2 + \left(\omega^2 - \bfk^2 \right) (-\dot{\theta}^2 + (\nabla \theta)^2 ) \right]^{1/2} \label{eqn:dispersion1}
\end{align}

\noindent where the $\pm$ refer to left-handed and right-handed helicity states, respectively. This \emph{birefringence} is a natural consequence of the pseudoscalar nature of the axion field.  It is worth emphasizing that the dispersion relation above does not assume that $\theta$ is a small parameter, only that its gradients are small in comparison to those of the Maxwell field. A large axion amplitude can lead to a tachyonic instability in one of the helicity modes \cite{jackiw}. Assuming spatial gradients of the axion field are subdominant compared to temporal gradients, the tachyonic instability occurs when $\dot{\theta} > |\bfk|$ which requires a central amplitude $a_0/f_a \gg {|\bfk|/( m_a \alpha_{_\text{EM}})} \gg 1$. Axion configurations with arbitrarily high central amplitudes can form in models with non-periodic potentials such as the ones considered in \cite{Eva2011}. However we restrict the following analysis to axion configurations wherein $\theta_0 < 1$. For more discussion of tachyonic instabilities in theories with pseudoscalar couplings to the Maxwell field, see \cite{jackiw}. Another peculiar feature of this dispersion relation is that it leads to a superluminal group velocity, which can raise questions of causality violation. Special relativity demands that no \emph{signal} can propagate faster than the speed of light in vacuum and in many situations group velocity is a poor proxy for signal velocity \cite{brillouin}, though there does not appear to be a generally accepted definition of the latter \cite{SuperluminalDebate}. The existence of superluminal group velocity has been demonstrated in a myriad of experiments. For an exhaustive list of these experiments and associated discussion, see \cite{MilonniSuperluminal}.



For light propagating in an axion background, the position, frequency, and wave number adjust so that (\ref{eqn:dispersion1}) is satisfied at each point along the path, defining geodesic trajectories \cite{WeinbergEikonal1962}. For a trajectory parameterized by an affine parameter $
\ell$, this leads to the condition

\begin{align}
    {d D^\pm \over d\ell} = {\partial D^\pm \over \partial x^\mu} {dx^\mu \over d \ell} + {\partial D^\pm \over \partial k^\mu} {dk^\mu \over d \ell} = 0
\end{align}

In analogy with Hamiltonian optics we then define trajectories such that $dx^\mu/d\ell = \partial D^\pm/\partial k_\mu, dk^\mu/d\ell = -\partial D^\pm/\partial x_\mu $. Using these definitions, we may use $t$ as our affine parameter, leading to the equations for the evolution of the position, momentum, and frequency along a trajectory.

\begin{subequations} \label{eqn:geodesics}
\begin{align}
    {d \bfx \over dt} &= {\partial D^\pm \over \partial \bfk} \left( {\partial D^\pm \over \partial \omega } \right)^{-1} \label{eqn:geodesica} \\
    {d \bfk \over dt} &= -{\partial D^\pm \over \partial \bfx} \left( {\partial D^\pm \over \partial \omega } \right)^{-1} \label{eqn:geodesicb} \\
    {d \omega \over dt} &= {\partial D^\pm \over \partial t} \left( {\partial D^\pm \over \partial \omega } \right)^{-1} \label{eqn:geodesicc}
\end{align}
\end{subequations}

\begin{figure*}
    \centering
    \includegraphics[scale=0.4]{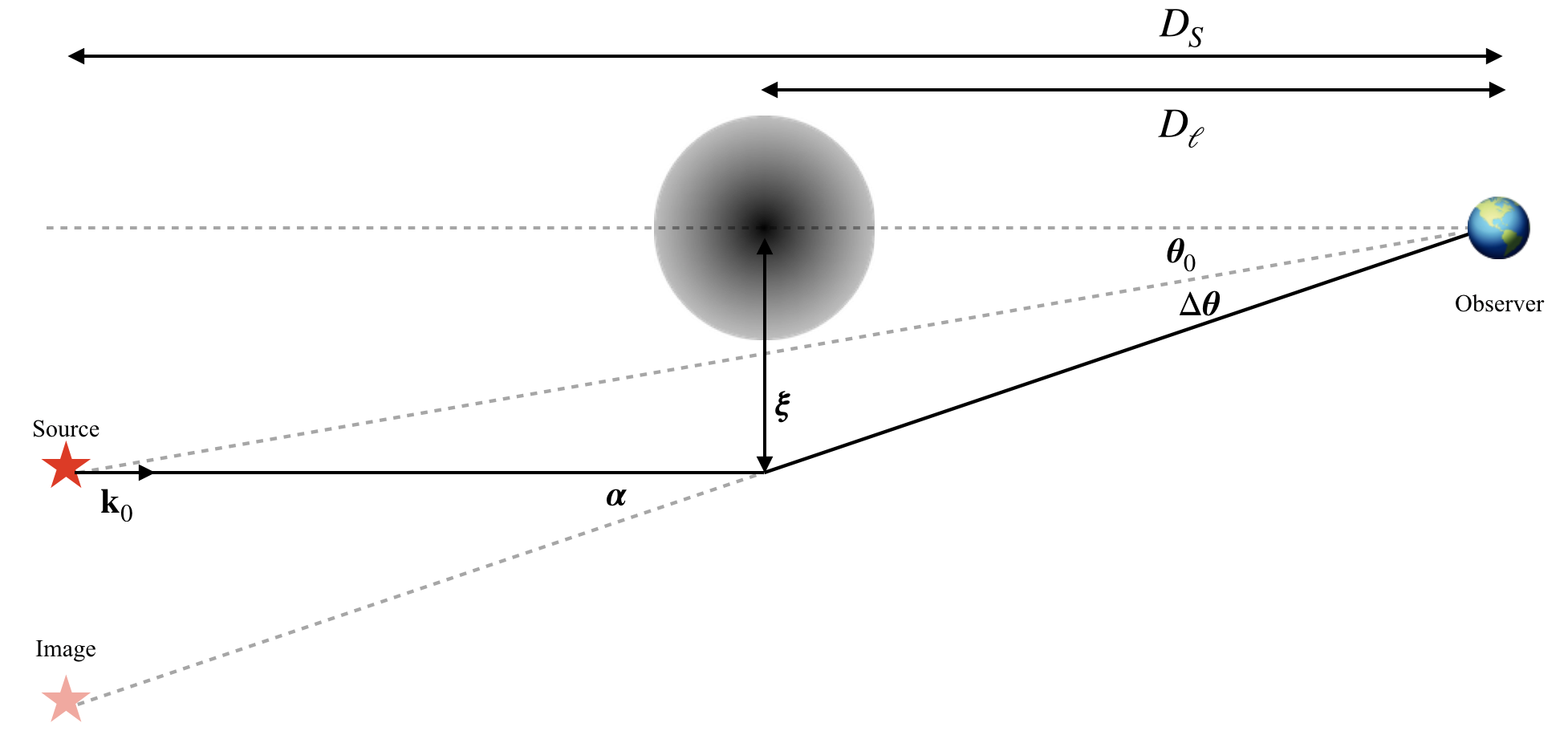}
    \caption{Geometry of light bending by axion stars. In this scenario, a source at distance $D_S$ and angular position on the sky (with respect to the observer) $\bftheta_0$ emits a light ray with momentum $\bfk_0$. The light ray intersects the lens plane at an impact parameter $\bfxi$ and gets deflected by an angle $\bfalpha$ towards the observer. We assume the  lens is at a distance $D_{_\ell} \ll D_S$ within the Milky Way. The angular separation between the image and the source is $\Delta \bftheta$. }
    \label{fig:lensingGeometry}
\end{figure*}

It appears from (\ref{eqn:geodesics}) that an intervening axion clump can imprint birefringent spectral distortions and momentum shifts (refraction) on background sources. Claims have been made that such effects are present at linear order in $\theta$ \cite{Plascencia_2018}. However, it was recently pointed out that in the presence of only an axion background, any spectral shifts are only sensitive to the axion configuration at the points of emission and observation \cite{blas2019chiral, mcdonald2019optical}. The authors of \cite{mcdonald2019optical, McDonald2020} claimed that the linear-order birefringent effect can be restored if ``medium'' effects were included. As an example, the authors discussed a cold plasma, though they claim any medium that endows the photon with a refractive index can restore the effect. We now evaluate this claim in the context of ``realistic'' axion clump models. As in Fig. \ref{fig:lensingGeometry}, we consider a light ray emitted from a source $S$ and passing by an axion clump with impact parameter $\bfxi$ and refracting towards the observer. To leading order, the deflection angle can be evaluated by integrating (\ref{eqn:geodesicb}) along the unperturbed ray path. This leads to a deflection angle, defined as $\bfalpha \equiv \delta\bfk/|\bfk|$ \cite{mcdonald2019optical}
\begin{widetext}
\begin{equation}
    \bfalpha(\bfxi) = \pm {1 \over 2 |\bfk_0|} \displaystyle\int_{t_{\text{em}}}^{t_{\text{obs}}} \left[ n_0 \nabla_\perp \dot{\theta}(t', \bfx_0(t';\bfxi)) + (\hat{\bfk}_0 \cdot \nabla) \nabla_\perp \theta(t', \bfx_0(t';\bfxi)) \right] dt' \label{eqn:bendinganglelinear}
\end{equation}
\end{widetext}

\noindent where $t_{\text{em}}$ and $t_{\text{obs}}$ are the times of emission and observation, $n_0$ is the refractive index of the background medium, $\nabla_\perp$ is the gradient transverse to the direction of propagation, $\bfx_0(t, \bfxi)$ is the unperturbed photon trajectory passing through an impact vector $\bfxi$ in the lens plane, and ${\bfk}_0$ is the wavenumber of the undeflected light ray. In the case of a cold plasma medium considered in \cite{mcdonald2019optical, McDonald2020}, $n_0 \approx 1 - {\omega_p^2/2|\bfk_0|^2}$ where $\omega_p$ is the plasma frequency and $\omega_0$ is the unperturbed photon frequency. As first noted by \cite{blas2019chiral}, in the absence of a background medium ($n_0 = 1$), the integrand of (\ref{eqn:bendinganglelinear}) is a total derivative, which makes the bending angle only sensitive to the boundary conditions. However, even in the presence of a medium, the effect may be washed out by the time oscillation of the axion field. Stable axion stars are usually non-relativistic (or semi-relativistic), as relativistic configurations enter the unstable region in the axion star phase space \cite{tkachev2018}. Stable, non-relativistic axion configurations are well-approximated by single harmonic with frequency of the axion mass. During a photon traversal, the axion undergoes a large number of oscillations per coherence length, effectively washing out the effect at linear order. The precise form of the suppression depends on the spatial profile of the axion clump. We emphasize that the arguments above regarding the bending of light apply to other effects such as spectral distortion as well. At quadratic order, however, light deflection is not averaged out as it traverses the axion configuration, though the effect is no longer birefringent. 

To compute the bending angle, we integrate terms in (\ref{eqn:geodesicb}) that are quadratic in $\theta$ over an unperturbed ray path. We will consider axion configurations with temporal gradients dominating spatial gradients, $\dot{\theta} \gg (\partial_i \theta)$. With this simplification, the deflection angle at quadratic order is approximately

\begin{align}
    \bfalpha(\bfxi) = {1 \over 8 |\bfk_0|^2} \displaystyle\int_{t_{\text{em}}}^{t_{\text{obs}}} \nabla_\perp \dot{\theta}^2(t', \bfx_0(t';\bfxi)) dt' \label{eqn:olas}
\end{align}

One feature of axion lensing that distinguishes it from gravitational lensing is the frequency dependence of the signal. By the equivalence principle, the gravitational bending angle is independent of frequency while, due to the dispersive nature of the axion medium, the optical lensing effect scales as the inverse square of the frequency. Thus a smoking gun signal of OLAS would be the presence of a lensing signal at low frequencies and the absence of a corresponding signal at higher frequencies.\footnote{Lensing effects by axion stars are very similar in nature to \emph{extreme scattering events}, thought to be caused by plasma inhomogeneities. Axion overdensities mimic the effects of plasma underdensities. However, the absence of coherent time oscillation of plasma lenses ensures that their effects are birefringent, distinguishing them from axion overdensities.} 

\section{Axion Clump Formation} \label{sec:clumpmodels}

We use the term ``axion clump'' to describe any inhomogeneous axion configuration with density larger than that predicted by $\Lambda$CDM. In this section, we discuss a few well-motivated axion clump models, classifying them by their formation mechanism.

\subsection{Axion Halos from Gravitational Collapse}

 Ultradense axion structures rely on the existence of large fluctuations in the axion field. These can form in one of two ways. If PQ symmetry is broken after inflation, causality demands that the axion field takes on different values in different Hubble patches. These values are drawn from a uniform distribution on the interval $[0, 2\pi]$, leading to $\mathcal{O}(1)$ fluctuations in the axion field on scales of order the horizon size at PQ symmetry breaking. At a scale $\Lambda_{\text{NP}}$, non-perturbative effects (e.g. instantons) generate a potential and hence a mass for the axion \cite{Gross1981}. In each Hubble patch the axion remains dynamically frozen at the value generated at PQ symmetry breaking until $H \sim m_a$ at which point the axion starts oscillating and generates $\mathcal{O}(1)$ isocurvature density fluctuations.\footnote{In many models, including the QCD axion, the axion mass has a nontrivial temperature dependence of the form $m_a = m_{a,0} (T/\Lambda_{\text{NP}})^{-n}$ where $n$ is a model-dependent index \cite{Gross1981}. For simplicity we consider $n=0$, but a more detailed analysis of QCD axions requires restoring this index.} These fluctuations will undergo gravitational collapse during radiation domination and subsequently form \emph{axion miniclusters} \cite{Hogan1988, KolbTkachev93, KolbTkachev941, KolbTkachev942}. Observational constraints on the tensor-to-scalar ratio provide an upper bound on the scale of inflation $H_I \lesssim 10^{14}$ GeV and hence on the axion decay constant in the minicluster scenario.\footnote{An intriguing scenario that circumvents this constraint occurs if the axion gets its mass from a strongly coupled hidden sector gauge theory with a first-order phase transition that takes approximately Hubble time to complete \cite{edhardy}.} The minicluster mass is given by the total amount of mass enclosed within a spherical volume of comoving radius $\pi/k_0$, where $k_0$ is the comoving horizon size when the axion starts oscillating. This leads to a minicluster mass

\begin{align}
    M_{\text{mc}} \approx 2.3 \times 10^{-13} M_\odot \left({10^{-6} \text{ eV } \over m_a} \right)^{3/2} \label{eqn:miniclustermass}
\end{align}

The typical radius of a minicluster is set by the (\ref{eqn:miniclustermass}) and the characteristic density 

\begin{align}
    \rho_{\text{mc}} \simeq 140 \Phi^3 (1 + \Phi) \rho_{a,\text{eq}}
\end{align}

\noindent where $\rho_{a,\text{eq}}$ is the average axion density at matter-radiation equality and $\Phi$ is the initial (dimensionless ) axion overdensity. Simulations give the best-fit cumulative mass fraction of miniclusters $f(\Phi > \Phi_0) = \left(1 + {\Phi_0 / a_1} \right)^{a_2}$ where $a_1 = 1.023$ and $a_2 = -0.462$  \cite{KolbTkachevPicoFemtolensing,Fairbairn2017}. A sizeable fraction of miniclusters are then predicted to have overdensities exceeding $\Phi = 100$. The corresponding minicluster radius is given by

\begin{align}
    R_{\text{mc}} = {1.8\times 10^7 \text{ km} \over \Phi^3 (1+\Phi)} \left({{10^{-6}} \text{ eV} \over m_a} \right)^{1/2}
\end{align}

Another progenitor theory for large axion density perturbations is the \emph{large-misalignment mechanism} \cite{marios2019}. If the axion is initially very close to the top of its potential, attractive self-interactions may delay the onset of oscillations. Once the axion starts oscillating, certain modes undergo parametric resonant growth sourced by the axion zero-mode and standard adiabatic curvature fluctuations, leading to denser, more numerous halos than predicted by $\Lambda$CDM. For a cosine potential, which is generic in theories with instanton contributions, such as the QCD axion, this requires an apparent tuning of the initial misalignment angle $a/f_a \approx \pi$. This apparent tuning can be circumvented by models that dynamically drive the misalignment angle to $\pi$ \cite{raymondco, yinwen2019}. The aforementioned parametric resonant growth is generic to potentials in which the axion has attractive self-interactions at large amplitude. 

For such large misalignment angles, modes with physical wavenumber $k_{\text{phys}} \sim H$ will go nonlinear earlier than in the $\Lambda$CDM scenario, leading to denser, \emph{large-misalignment halos}. The large-misalignment mechanism produces dense axion halos with mass peaked around \cite{marios2019}

\begin{align}
    M_* = {4\pi \over 3} \left( {\lambda_* \over 2} \right)^3 \rho_{_{\text{DM}, 0}} = 5 \times 10^{-15} M_\odot \left({10^{-6} \text{ eV} \over m_a} \right)^{3/2} \label{eqn:femtomass}
\end{align}

\noindent where $\lambda_* = 2\pi/\sqrt{2 m_a H_{\text{eq}} a_{\text{eq}}^2}$ (here ``eq'' corresponds to values at matter-radiation equality) and $\rho_{_{\text{DM}, 0}}$ is the average dark matter density at present. The densities of these objects are parametrically larger than those of CDM halos by a factor $\mathcal{B}$ which is taken to be in the range $10^3 \lesssim \mathcal{B} \lesssim 10^7$. The typical size of these objects is determined by the scale radius (cf. the discussion after (\ref{eqn:NFW}))

\begin{align} 
    R_s \approx 4\times 10^{-6} \text{ pc} \left({10^{-6} \text{ eV} \over m_a }\right)^{1/2} \left( {10^3 \over \mathcal{B}} \right)^{1/3} \label{eqn:femtoradius}
\end{align}

In addition to the modes discussed above, certain modes can go nonlinear deep into radiation-domination and collapse due to self-interactions, leading to extremely dense objects called \emph{oscillons}. We will discuss these objects in the following section. 

\vspace{7mm}

\subsection{Solitons and Oscillons}

Gravitational collapse of axion halos is eventually halted by the repulsive kinetic (uncertainty) pressure. This occurs when the size of the halo becomes comparable to the de Broglie wavelength of the axion field. The resulting ground state configuration is called a \emph{soliton} \cite{Khlebnikov1,Khlebnikov2, chavanis1, chavanis2, GuthBEC}. Solitons are similar to the \emph{bose stars} described in \cite{Kaup1968,Ruffini69}. Stars composed of bosons of mass $m$ were shown to be unstable to collapse above a critical mass $M^*_{\text{bose}} \simeq 0.6\mpl^2/m$ \cite{BREIT1984329, UrenaLopez2002, Barranco2013}. When the constituent bosons are axions, the resultant objects are called \emph{axion stars}. At masses much less than the critical mass the properties of bose stars and axion stars coincide. As the mass approaches the critical value, weak, attractive axion self-interactions can play a role in the balance of forces. Axion stars have a critical mass (and corresponding critical central amplitude) given by 

\begin{align}
    M_{\text{AS}}^* &\simeq 10 {f_a \mpl \over m_a} \label{eqn:mcrit} \\
    \phi_0^* &\simeq {10.5 \over |\lambda_4|} \left( {f_a \over \mpl} \right) \simeq {8.6 \times 10^{-7} \over |\lambda_4|} \left( {f_a \over 10^{12} \text{ GeV}} \right)
\end{align}

\noindent where $\phi_0^* = a_0^*/f_a$ and $\lambda_4$ is an $\mathcal{O}(1)$ dimensionless coupling constant associated with the quartic term in the potential. For the QCD axion with a chiral potential, $\lambda_4 = -0.34$. Axions stars with mass less than this value are called \emph{dilute axion stars} \cite{baum2017}. They have a well-defined mass-radius relationship \cite{Bernal2011}

\begin{align}
    R_{99} &\simeq {9.95 \mpl^2 \over M m_a^2}
\end{align}

\noindent where $M$ is the star mass, $R_{99}$ is the radius containing 99\% of the mass. The spatial profile will be important for understanding the lifetime and lensing effects, however, there is no consensus on its form in the literature. The profile is often assumed to be well-approximated by a Gaussian in, for example, \cite{chavanis1,chavanis2, marios2019} and an exponential ($ \rho(r) \propto \exp(-r/R) $) in \cite{iwazaki2014, Iwazaki2015, bai2018}. Although dedicated numerical studies of the spatial profiles of bosons stars presented in \cite{Kling1, Kling2} favor an exponential profile over a Gaussian, however we conservatively adopt the latter. 

Solitons with mass parametrically lower than the critical mass by a factor of roughly $(H_{\text{eq}}/m_a)^{1/4}$ can be formed directly from the large-misalignment mechanism \cite{marios2019}. These solitons can either merge with each other or accrete mass from their surrounding halos through the gravitational cooling mechanism, resulting in a present-day abundance of near-critical solitons. Gravitational cooling can also lead to the formation of solitons in the minicluster scenario as shown in \cite{tkachev2018}.

When the axion amplitude is greater than $a_0^*$ but less than unity, the resulting \emph{critical axion stars} are unstable to any perturbations that increase the central amplitude \cite{chavanis2}. As the amplitude becomes $\mathcal{O}(1)$, quartic self-interactions are no longer sufficient to describe the balance of forces. Considering the full potential leads to metastable objects known as \emph{oscillons} or \emph{dense axion stars} \cite{braatenDense2016, baum2017, chavanis3}. Oscillons may be formed as the by-product of collapse of super-critical solitons. Solitons may become supercritical through mergers of or accretion by near-critical solitons \cite{MustafaClustering}. As mentioned earlier, oscillons may also form directly from the large-misalignment mechanism when axion fluctuations collapse due to self-interactions deep in radiation-domination \cite{marios2019}.

As mentioned earlier, oscillons are metastable states wherein the outward kinetic pressure is balanced by self-interactions described by the full potential. We quote a few oscillon features present in a wide variety of initial conditions of simulations performed in \cite{marios2019}. Firstly, the central density is of order the natural scale, $m_a^2 f_a^2$, present in the cosine potential and in more exotic potentials like the ones in \cite{Eva2011,OscillonDarkMatter}. In addition, the size of these objects is roughly $R =  n/m_a$ where $n \sim \mathcal{O}(10)$. Assuming a roughly flat internal density profile, we can estimate the oscillon mass to be $M_\text{osc} \sim n^3 f_a^2/m_a$. 

Another property of oscillons is that general solutions involve infinitely many harmonics $\omega_0, 3 \omega_0, 5 \omega_0, \dots$. Here $\omega_0$ is the fundamental frequency, which is roughly $\omega \approx 0.9 m_a$, because of the small binding energy per axion. Numerical studies \cite{baum2017, marios2019} have shown that the fundamental harmonic is dominant. Motivated by these studies, we work in the single-harmonic approximation. The spectrum of emitted axion radiation indicates that oscillons may be metastable due to number-changing processes. The metastability of oscillons plays an important role in their viability as a DM candidate.
 
\subsection{Oscillons as Dark Matter Candidates} \label{sec:lifetime}

Crucial to the observation of oscillons is their present-day abundance. This quantity depends on the formation mechanism and lifetime. While solitons are expected to have lifetimes greater than the age of the universe \cite{EbyLifetime,EbyDecayCondensate}, the oscillon lifetime is not well understood. Since oscillons probe the full potential they are metastable to decay through number-changing processes such as $3a \rightarrow a $, $4a \rightarrow 2a$, etc. By conservation of energy, the outgoing axions will be relativistic for most number-changing processes. Thus, the oscillon lifetime is very sensitive to its size and spatial profile as those determine the suppression of the rate of production of high-momentum axion modes. 

For the cosine potential, oscillons decay in time $\tau_{\text{osc}} \sim 10^3/{m_a}$, which means that they are unlikely to survive past matter-radiation equality assuming they were formed via the large-misalignment mechanism. For flatter potentials oscillons have been observed, in numerical simulations, to have lifetimes $\tau \gtrsim 10^8/m_a$ \cite{OscillonDarkMatter}, with no apparent upper bound. Similarly, arguments by \cite{EbyLifetime} suggest that for large oscillons with certain spatial profiles the rates of number-changing processes can be exponentially suppressed leading to large lifetimes. Mechanism to naturally generate flat potentials were proposed in \cite{Eva2011, GengFlatten}. We consider a general class of well-motivated potentials \cite{Dubovsky_2012},

\begin{align}
    V(a) &= {m^2 f_a^2 \over 2 p} \left[ \left(1 + {a^2 \over f_a^2} \right)^p - 1 \right] \label{eqn:potential}
\end{align}

\noindent where $0 \le p \le 1 $ gives axion monodromy-like potentials \cite{Eva2008, EvaMcAllister2008, MustafaOscillon} and $p<0$ gives potentials that flatten at large field values. So-called \emph{plateau} potentials have been discussed in an inflationary context in \cite{LozanovOscillon, LindeKallosh}. Potentials with $p < 1$ can form potentially long-lived oscillons  \cite{OscillonDarkMatter, MustafaOscillonDecay}. Monodromy-like potentials were shown to lead to lifetimes in the range $(10^8-10^9)/m_a$, which are considerable for very low-mass axions. However, the mass range probed by OLAS is $10^{-9}$ eV $\lesssim m_a \lesssim 10^{-4}$ eV. The mass range is set by the frequencies observed by radio telescopes. Thus it seems unlikely that these potentials can lead to a present-day abundance of oscillons. For $p<0$, where potentials become asymptotically flat, oscillons were shown to live for $\tau > 10^9/m_a$ with no sign of decay. It is possible that oscillons formed from such potentials can be present today and form an $\mathcal{O}(1)$ fraction of DM. Therefore in the following discussion of observational signatures, we assume oscillons formed not from QCD axions, but from axions with potential (\ref{eqn:potential}), with $p=-1/2$. Detailed simulations of oscillon properties may be found in \cite{Kawasaki2020, Cyncynates2021}. 

\section{Observational Prospects} \label{sec:signatures}

In Sec. \ref{sec:AEM} we showed that inhomogeneous axion configurations can lead to deflection of light rays. This is analogous to the deflection of light due to massive objects in general relativity, with some key differences. We thus expect much of the gravitational lensing formalism to apply to the case of optical lensing by axion stars.\footnote{We emphasize that this effect is analogous to but distinct from gravitational lensing. Proposals to search for axion clumps using gravitational lensing include \cite{Fairbairn2017, fairbairn2018microlensing, marios2019,KolbTkachevPicoFemtolensing,kenAstrometry}} OLAS can produce multiple images as well as magnification of background sources that could potentially be detected by strong lensing and microlensing searches. We will investigate these signatures in \cite{prabhu2}. Even in the absence of such dramatic effects, axion clumps may induce apparent anomalous motion of background sources during close line-of-sight approaches. Astrometric missions such as \emph{Gaia} \cite{Gaia2016} seek to measure the positions, velocities, and distances of background sources with unprecedented precision. At lower frequencies, the upcoming Square Kilometer Array (SKA) is slated to be the largest and most sensitive radio interferometer to date. The array will be built in two phases. The first phase, called SKA1, which will use only about 10\% of the collecting area of SKA2, is expected to be deployed in the early 2020s. SKA is projected to continuously cover frequencies between 50 MHz and 14 GHz in its first two phases and frequencies above 30 GHz in its third phase. One of the many astrometric capabilities of SKA is its ability to measure distances and proper motions of background radio sources with unprecedented angular precision \cite{fomalont}. The angular displacement of a source is often difficult to determine from a single measurement due to the intrinsic uncertainty in the unlensed position of the source. In this section, we calculate the sensitivity of certain time-domain astrometric observables outlined in \cite{kenAstrometry} to OLAS. In order to confirm that a lensing event is due to OLAS and not gravitational lensing simultaneous measurements must be made at different frequencies. If two detectors observing in different frequency ranges record different positional shifts of the same background source then the two observations could constitute a signal of OLAS. For example, if Gaia, operating in the optical range, and SKA, operating in the radio range, record different positional shifts of the same background source, with the frequency-dependence of the positional shifts described by (\ref{eqn:olas}) then we can conclude that the differential positional shift is due to OLAS and not gravitational lensing. We note that multi-wavelength observations of hundreds of quasars have already been made with Gaia and Very Large Baseline Interferometry (VLBI) \cite{VLBIGaia2012}. 

For a wide range of axion masses, the corresponding solitons and oscillons are approximately point-like and traverse large arcs on the sky during the duration of the mission (assumed to be $\tau = 5$ years). The observable effects of the axion star can be divided into two categories \footnote{When the proper motion of a lens is smaller than the typical minimum impact parameter in a source-lens pair, there will be anomalously high velocities and accelerations of the sources. These are called \emph{outliers}. Since the axion-photon interaction is short-range in comparison to gravity these observables are not well-suited to oscillon searches using optical lensing}:

\begin{itemize}
    \item \textbf{Rare Single Events:} Over the duration of the mission, we expect there to be a very close approach (in the lens plane) between a lens and a background source. In this close approach the source will appear to traverse a nonlinear trajectory before returning to its original position. This observable is called a \emph{mono-blip}.
    \item \textbf{Multi-Source Observables:} If the angular distance traveled by the lens is larger than the angular separation between sources, there will be correlated anomalous motion of stars, called a \emph{multi-blip} signal. 
\end{itemize}

\subsection{Blip Sensitivity}

Blips are the most suitable observables when the distance traversed by the lens during the mission is larger than the size of the lens and the minimum expected approach between a source-lens pair. For the axion clump models and masses of interest, the above condition will usually apply. To compute the sensitivity of an SKA astrometric mission to blips caused by axion stars, we propose making high-cadence, all-sky measurements of the positions of many background radio sources. In the absence of any lenses, the background sources remain approximately fixed on the sky. This is expected since the proper angular motion of radio sources at cosmological distances over the course of the 5 year mission is much less than the angular precision expected for SKA, $\Delta \theta \simeq 10 \ \mu$as \cite{fomalont}. In the presence of a lens with trajectory $\bfx_{_\ell}(t)$ (in the lens plane), OLAS predicts that a source at angular position $\bftheta_{i,0}$ will undergo an apparent non-linear lensing trajectory, $\bftheta_i(t)$, given by

\begin{align}
    \bftheta_i(t) = \bftheta_{i,0} + \left(1 - {D_{_\ell} \over D_S} \right) \bfalpha\left[D_{_\ell} \bftheta_{i,0} - \bfx_{_\ell}(t) \right] \label{eqn:angulardisplacement}
\end{align}

\noindent where $D_{_\ell}$ and $D_S$ are the distance to the lens and source, respectively and $\bfalpha$ is the deflection angle, computed in (\ref{eqn:olas}). The term in square brackets is the impact parameter at which light from the source passes by the lens. We assume galactic lenses and radio
sources that are at cosmological distances. Therefore, we can neglect the term $D_{_\ell}/D_S$. To test whether the presence of an axion star lens is consistent with the data, we construct a test statistic, $\mathcal{B}$ , that encodes the overlap of the observed angular position residuals with the angular position shift predicted by OLAS (\ref{eqn:angulardisplacement}), weighted by the angular precision for each measurement, $\sigma_{\delta\theta_i}$ \cite{kenAstrometry}. If the lens trajectory used to compute $\bftheta_i(t)$ is correct, the expected angular position residuals should coincide with the predictions made by OLAS. The expectation value of the test-statistic can then be calculated to be \cite{kenAstrometry}

\begin{widetext}
\begin{align}
    \left\langle \mathcal{B}[\bfx_{_\ell}(t)]\right\rangle &= \left\langle \displaystyle\sum_{i = 1 }^N {1 \over \sigma^2_{\delta \theta_i}} \displaystyle\sum_{j = 1}^n \alpha^2[\bfx_\ell(t_n)] \right \rangle \simeq \left({\gagg^2 m_a^2 a_0^2 \over  8 \sigma_{_{\delta\theta,\text{eff}}} \omega^2}\right)^2 \left\langle \displaystyle\sum_{i = 1}^N {f_{\text{rep}} \over v_{_\ell} }\displaystyle\int_{-\infty}^\infty f^2\left( \sqrt{b_{i\ell}^2 + x^2} \right) dx \right \rangle  \label{eqn:teststatistic}
\end{align}
\end{widetext}

\noindent where the first sum is performed over the $N$ sources and the second sum over the $n$ observation times, $\omega$ is the frequency of light, $v_{_\ell}$ is the velocity of the axion lens, $\sigma^2_{\delta \theta_i}$ is the angular position noise per observation, $b_{i \ell}$ is the minimum expected impact parameter between a given source $i$ and lens $\ell$, and $f(x)$ is the profile-dependent form factor defined as

\begin{align} 
f(\bfxi) &\simeq \displaystyle\int_{-\infty}^\infty \left.(\nabla_{\perp}\bar{\theta}^2(\bfx, t')) \right|_{\bfx = \bfx_0(t'; \bfxi)} dt'
\end{align}

\noindent where $\bar{\theta}(\bfx, t) = \theta(\bfx, t)/\theta_0$ and $\bfx_0(t; \bfxi)$ is the unperturbed path of the light ray with impact parameter $\bfxi$, and $\nabla_\perp$ refers to the gradient normal to the direction of propagation. The second step in (\ref{eqn:teststatistic}) involved approximating the sum over observation times as a continuous integral with integration variable $x$. For a given lens path, the signal-to-noise ratio will be given by $\sqrt{ \left \langle \mathcal{B}[\bfx_\ell(t)]\right\rangle}$. This can be large due to a single lensing event with very low impact parameter (mono-blip) or a lens path that passes close to several sources $\sum_{\text{sources}} \gg 1$ (multi-blip). We will find that for the axion clump models of interest, the former possibility is best-suited for detection. 

For mono-blip signals, we will be interested in the closest expected approach, $b_{i\ell}$, between any galactic lens and photons from any source on the sky during the duration of the mission. This can be approximated as 

\begin{align}
    \left \langle\min\limits_{i,\ell} b_{i \ell} \right \rangle &= {M_{_\ell} \over v_{_\ell} \tau \rho_{_\ell} D_S \Sigma_0 \Delta \Omega} \label{eqn:minbil}
\end{align}

\noindent where $M_{_\ell}$ and $v_{_\ell}$ are the mass and velocity of the lens, $\tau$ is the duration of the mission, $\rho_{_\ell}$ is the average density of lenses in the region of interest (not to be confused with the density of a single lens), $D_S$ is the distance to the sources, $\Sigma_0$ is the angular number density of sources, and $\Delta \Omega$ is the angular area covered by the survey. We now discuss the astrometric sensitivity of SKA to various axion clump models in the OLAS scenario. We split the consideration into two cases: (1) ``diffuse'' objects (miniclusters and large-misalignment halos) and (2) compact objects (solitons and oscillons).

\subsection{Minicluster and Large-Misalignment Halos}

The formation of axion miniclusters and large-misalignment halos is analogous to that of $\Lambda$CDM halos. We therefore assume they have the same density profile. In $\Lambda$CDM, $N$-body simulations of gravitational collapse of overdensities show the formation of virialized halos that are well-fitted by a Navarro-Frenk-White (NFW) profile over a broad range of initial conditions \cite{NFW}

\begin{align}
    \rho_{_\text{NFW}}(r) &= {4 \rho_s \over {r \over R_s} \left( 1 + {r \over R_s} \right)^2} \label{eqn:NFW}
\end{align}

\noindent where $\rho_s $ is the scale density, the characteristic density of the halo, and $R_s$ is the scale radius, the distance at which the halo density is the scale density. The amount of mass enclosed in the scale radius is called the scale mass. An important property of this profile is the apparent ``cuspy'' behavior at small radii. This is in conflict with the ``cored'' DM density profile inferred from rotation curves of dwarf galaxies \cite{CuspCore94,CuspCore2015}. This conflict is known as the \emph{core-cusp problem}, which may be resolved by including baryonic feedback or the formation of solitonic cores at the center of CDM halos. This problem highlights the fact that the NFW profile may not apply when $r \ll R_s$. 

The mono-blip test statistic for an NFW profile is well-approximated by

\begin{widetext}
\begin{align}
    \left\langle \mathcal{B}_{\text{NFW}}[\bfx_{_\ell}(t)]\right\rangle \simeq \left( {\gagg^2 \rho_s \over 4 \sigma_{_{\delta\theta,\text{eff}}}  \omega^2}  \right) \sqrt{{ \pi f_{\text{rep}} R_s \over v_{_\ell}} \left( \left \langle\min\limits_{i,\ell} b_{i \ell} \right \rangle \over R_s  \right)^{-1}}
\end{align}
\end{widetext}

Based on (\ref{eqn:miniclustermass})-(\ref{eqn:femtoradius}), we find that for blip searches, sensitivity to miniclusters and large-misalignment halos relies on impact parameters much smaller than the scale radius. The radii involved are lower than the spatial resolution of $N$-body simulations and thus it is unclear if the NFW profile can be trusted in this regime. The situation in which a solitonic core is formed at the center of the halo is considered in the following subsection. The argument above also applies to the case of multi-blip searches. Even the $\sqrt{N}$ enhancement from correlated anomalous motion of many stars will be insufficient to make OLAS observable. Searches for blip signals sourced by gravitational lensing is a more promising avenue for these objects \cite{kenAstrometry}. 

\subsection{Solitons and Oscillons}

For various profiles, the deflection of light by axion stars is maximized when the impact parameter is of order of the radius of the object. Since solitons and oscillons are very small and compact, we expect mono-blip searches to be the most promising. For a Gaussian profile, the mono-blip SNR is given by

\begin{widetext}
\begin{align}
    \text{SNR}^{\text{mono}}_{\text{Gauss}} \equiv \left \langle \mathcal{B}\left[ \bfx_{_\ell}(t) \right] \right \rangle^{1/2} \simeq \left({\gagg^2 m_a^2 a_0^2 \over 8 \sigma_{_{\delta\theta,\text{eff}}}  \omega^2} \right) e^{ - {\left \langle\min\limits_{i,\ell} b_{i \ell} \right \rangle^2/ R^2} } \left[ {f_{\text{rep}} R \over v_{_\ell}} \times \left( 1 + {4 \left \langle\min\limits_{i,\ell} b_{i \ell} \right \rangle^2 \over R^2} \right) \right]^{1/2} \label{eqn:snrmonogauss}
\end{align}
\end{widetext}

The radius containing 99\% of the energy of the star, $R_{99}$ is related to $R$ by a potential and profile-dependent $\mathcal{O}(1)$ constant of proportionality. We compare (\ref{eqn:snrmonogauss}) to the mono-blip SNR from gravitational lenses \cite{kenAstrometry}

\begin{align}
    \text{SNR}^{\text{mono}}_{G} \simeq {4 G M_{_\ell} \over \sigma_{_{\delta\theta,\text{eff}}}} \sqrt{{\pi f_{\text{rep}} \over v_{_\ell}} {1 \over \left \langle\min\limits_{i,\ell} b_{i \ell} \right \rangle } }
\end{align}

\noindent where $M_{_\ell}$ is the mass of the lens, and $\left\langle\min\limits_{i,\ell} b_{i \ell} \right \rangle$ is given in (\ref{eqn:minbil}). In Fig. \ref{fig:OscillonReach} the mono-blip sensitivity projections of OLAS and gravitational lensing are shown. Since gravitational lensing is not frequency dependent, we use the parameters expected for \emph{Gaia} sampling of disk stars: $\sigma_{_{\delta\theta,\text{eff}}} = 100  \ \mu$as, $f_{\text{rep}} = 14/$day, $D_S = 10$ kpc, $\tau = 5$ years, $\Sigma_0 = 4.6 \times 10^9$ rad$^{-2}$, and $\Delta \Omega = 0.2$ in Fig. \ref{fig:OscillonReach}. 

\subsubsection{Solitons}

In this section we discuss the projected sensitivity of our proposal to solitons. Since the stability of solitons is not sensitive to the choice of axion potential we consider solitons composed of QCD axions. In conventional models of the QCD axion the coupling of axions to photons is assumed to lie in a narrow band in parameter space. However, there exist several mechanisms to expand the allowed range of couplings. In particular, many models predict axion-photon couplings much larger than those in standard KSVZ \cite{KimOG, SVZOG} and DFSZ \cite{DFSOG,ZOG} realizations of the QCD axion \cite{photophilic2016,DiLuzio2017,DiLuzio20172}. For a list of mechanisms that can generate large axion-photon couplings, see \cite{Prateek2018}. Based on the arguments above, we assume the axion-photon coupling is a free parameter that is independent of $f_a$. 

\begin{figure*}
    \centering
    \includegraphics[scale=0.5]{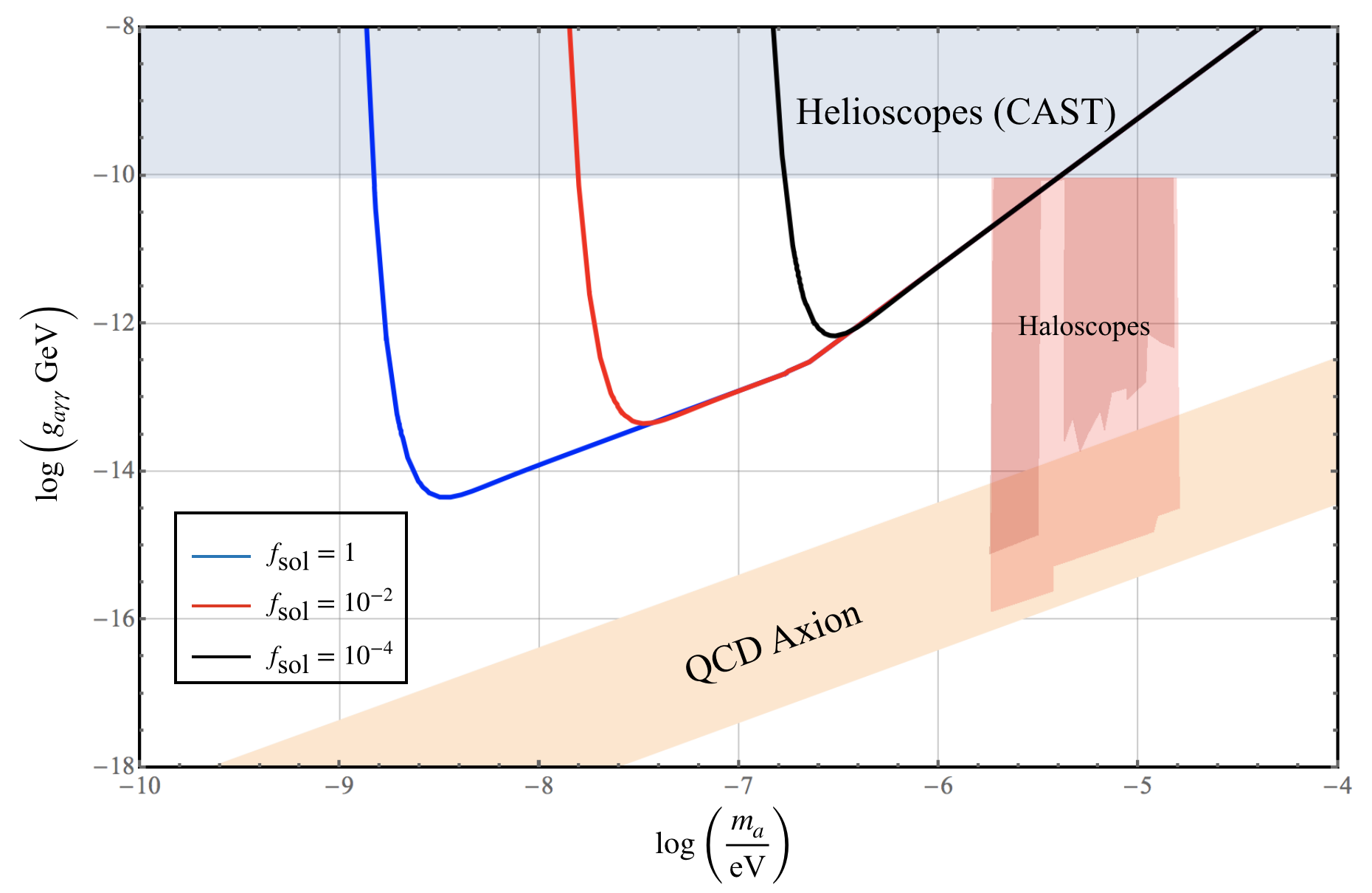}
    \caption{Projected sensitivity of an SKA mono-blip search to critical-mass solitons with fractional abundance $f_{\text{sol}} = 1$ (blue), $f_\text{sol} = 10^{-2}$ (red), $f_\text{sol} = 10^{-4}$ (black) where $f_\text{sol} \equiv \rho_{\text{sol}}/\rho_{_\text{DM}}$.  We have assumed an angular number density $\Sigma_0 = 10^7$ rad$^{-2}$ of background radio sources covering the entire sky, a mission time of $\tau = 5$ years, lens velocity $v_{_\ell} = 10^{-3}$, an observing cadence of $f_{\text{rep}} = 1/$day, and an angular precision of $\sigma_{_{\delta\theta,\text{eff}}} = 10 \ \mu$as. }
    \label{fig:SolitonReach}
\end{figure*}

Constraining the fractional abundance of soliton DM requires knowing the mass function for these objects. Using the Press-Schechter formalism \cite{PressSchechter}, the authors of \cite{fairbairn2018microlensing} calculate the mass function for axion miniclusters. Such a computation for solitons requires determining the formation mechanism and the merger and accretion rates, which is beyond the scope of this work. We investigate the possibility of detecting solitons with OLAS by assuming a Dirac-delta function mass function, centered at the critical mass (\ref{eqn:mcrit}). The projected sensitivity is shown in Fig. \ref{fig:SolitonReach}, assuming the solitons are described by a Gaussian density profile. At low axion mass gradients of the axion field are suppressed, leading to a smaller deflection angle. 

For critical solitons, we only have sensitivity if the axion model has significant enhancements in the axion-photon coupling compared to conventional QCD axion models. This may be realized in the models discussed at the beginning of the section. However, at such high couplings critical solitons may be unstable to parametric resonant decay into photons \cite{Hertzberg_2018, levkov2020radioemission, Hertzberg_2020, Amin2020vja, Amin2021tnq}. 

\subsubsection{Oscillons}

\begin{figure*}
    \centering
    \includegraphics[scale=0.5]{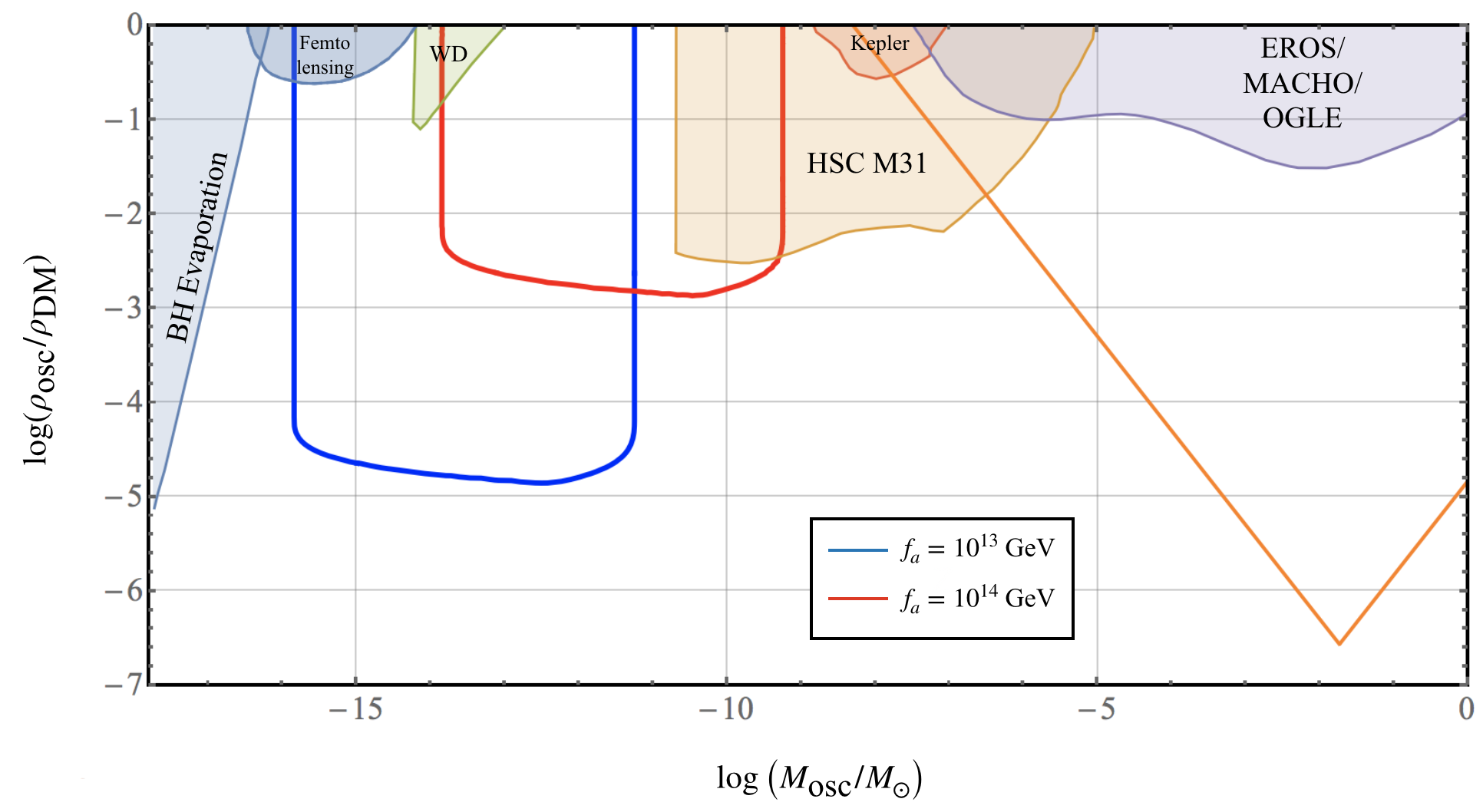}
    \caption{Estimated sensitivity of SKA mono-blip searches to optical lensing by oscillons. The plot shows the projected sensitivity to oscillons with $f_a = 10^{13}$ GeV (blue) and $10^{14}$ GeV (red). Estimates are calculated for central axion density $\phi_c = 5 f_a$, and axion-photon coupling $\gagg = \alpha/f_a$. For both plots, we have assumed an angular number density $\Sigma_0 = 10^7$ rad$^{-2}$ of background radio sources covering the entire sky, a mission time of $\tau = 5$ years, lens velocity $v_{_\ell} = 10^{-3}$, an observing cadence of $f_{\text{rep}} = 1/$day, and an angular precision of $\sigma_{_{\delta\theta,\text{eff}}} = 10 \ \mu$as. Existing observational constraints come from the absence of extragalactic $\gamma$-rays from PBH evaporation \cite{PBHEvaporation}, femtolensing of $\gamma$-ray bursts \cite{femtolensing2012}, the present-day abundance of white dwarfs \cite{PBHWD2015}, and microlensing from the Kepler mission \cite{Kepler2014}, the EROS/MACHO mission \cite{EROSMACHO2007}, and Subaru/HSC Andromeda observations \cite{HSCSubaru2017}. The thick orange lines represents the projected sensitivity of  \emph{Gaia} monoblip searches of the galactic disk using the gravitational deflection of light \cite{kenAstrometry}.}
    \label{fig:OscillonReach}
\end{figure*}

We also consider the possible lensing signatures of oscillons. As discussed in Sec. \ref{sec:lifetime}, such signatures are contingent upon oscillons being long-lived on cosmological timescales. Thus the following discussion does not apply to oscillons formed from a cosine potential. For non-periodic potentials, the central amplitude can theoretically be arbitrarily large, although the results of Sec. \ref{sec:AEM} rely on the assumption that $\gagg a_0 < 1$. Assuming $\gagg \sim \alpha/f_a$ yields the constraint $a_0/f_a \lesssim 10^{2}$. The reach of SKA mono-blip searches for optical lensing by oscillons is shown in Fig. \ref{fig:OscillonReach}. For fixed $f_a$, the oscillon mass range probed is determined by the axion mass. Since light-bending is proportional to gradients of the axion field, it is suppressed at low axion mass (high oscillon mass). The abrupt loss of sensitivity at high axion mass (low oscillon mass) is an artifact of the breakdown of the eikonal approximation that occurs when the axion mass becomes comparable to the highest frequency observed by SKA. Extending the sensitivity to lower masses can be achieved numerically in the regime where the eikonal approximation is no longer valid. We find that for $f_a \lesssim 10^{15}$ GeV, this search will be sensitive to oscillon masses in a range unconstrained by existing observational constraints. This appears to be true over a variety of well-motivated spatial profiles and central axion densities. 

Existing and proposed constraints on the abundance of general astrophysical objects come from gravitational microlensing surveys performed by MACHO \cite{MACHO2001}, EROS \cite{EROSMACHO2007}, OGLE \cite{OGLE2011}, HSC/Subaru \cite{HSCSubaru2017}, on objects with mass exceeding $10^{-11} \ M_\odot$. As mentioned in Sec. \ref{sec:intro}, constraints on general objects of mass less then $10^{-11} M_\odot$ are scarce. One proposal to probe masses in this range was femtolensing of gamma-ray bursts (GRBs); the proposal has, however, been consequently weakened by the inclusion of finite source size and wave effects \cite{Katz_2018}. In Fig. \ref{fig:OscillonReach}, we have also included constraints on PBH masses, although they do not apply to oscillons or other axion stars. Constraints on PBHs arise as a requirement that the time over which they dissipate to Hawking radiation be greater than the age of the Universe and the requirement that the Hawking radiation does not result in an excess in the gamma-ray background \cite{PBHBBN2010}. Finally, PBHs with masses between $5 \times 10^{-15} M_\odot$ and $5 \times 10^{-14} M_\odot$ can heat white dwarfs (WD) through dynamical friction. This can lead to runaway thermonuclear fusion in and subsequent explosion of the WDs. The observed distribution of WDs constrains PBHs in the aforementioned mass range \cite{PBHWD2015}.

\section{Conclusions \label{sec:conclusions}}

In this paper, we have shown that light rays can be deflected by inhomogeneous axion configurations due to axion-photon interactions. We emphasize that this effect is distinct from and in some cases more significant than the canonical bending of light due to gravity. We addressed some of the previous claims of the possibility of birefringent light bending that occurs at linear order in the axion-photon coupling and argued that in realistic axion clump models, an effect is only present at quadratic order, even in the presence of a background refractive medium. We discussed the sensitivity of SKA to the detection of anomalous shifts in the apparent positions of background radio sources due to optical lensing by axion stars. This complements existing proposals to observe the astrometric effects of dark lenses through gravitational lensing, but probes a completely different region of parameter space. For near critical-mass solitons, the absence of frequency-dependent blip signals in SKA can place constraints on the axion-photon coupling in the mass range $10^{-9} \text{ eV} \lesssim m_a \lesssim 10^{-4} \text{ eV}$, even assuming solitons make up a relatively small fraction of DM. However, we only find sensitivity to QCD axion models with enhanced axion-photon couplings, where instability to photon perturbations may dominate. We expect sensitivity to oscillons over a wide region of parameter space. It is interesting to note that the oscillon masses probed have significant overlap with the region of parameter space $M \in [10^{-14} M_\odot, 10^{-11} M_\odot]$, where there exist no robust bounds from gravitational microlensing. Optical lensing by axion stars could provide the first  model-dependent bounds on compact objects in this mass range. These sensitivity projections rely on oscillons being long-lived over cosmological time-scales. While this is unlikely to be the case for the QCD axion, it may be true in well-motivated models with flat potentials. For sufficiently dense axion clumps, optical lensing may give rise to dramatic lensing signatures such as significant magnification and the production of multiple images. These signatures may be enhanced when the clump is in the vicinity of a critical curve of another massive object. These effects and their observational prospects will be discussed in future work.

\begin{acknowledgments}
The author is grateful to Savas Dimopoulos, Davide Racco, and Mustafa Amin for enlightening discussions during various stages of this work and for useful comments on preliminary drafts of this manuscript. This work was supported by the National Science Foundation under Grant No.
PHYS- 1720397 and the Gordon and Betty Moore Foundation Grant GBMF7946. The author
acknowledges the support of the Fletcher Jones Foundation and the National Science Foundation (NSF) Graduate Research Fellowship Program. 
\end{acknowledgments}

\appendix

\bibliography{OLAS1v1}

\providecommand{\noopsort}[1]{}\providecommand{\singleletter}[1]{#1}%
\begin{thebibliography}{103}%
\makeatletter
\providecommand \@ifxundefined [1]{%
 \@ifx{#1\undefined}
}%
\providecommand \@ifnum [1]{%
 \ifnum #1\expandafter \@firstoftwo
 \else \expandafter \@secondoftwo
 \fi
}%
\providecommand \@ifx [1]{%
 \ifx #1\expandafter \@firstoftwo
 \else \expandafter \@secondoftwo
 \fi
}%
\providecommand \natexlab [1]{#1}%
\providecommand \enquote  [1]{``#1''}%
\providecommand \bibnamefont  [1]{#1}%
\providecommand \bibfnamefont [1]{#1}%
\providecommand \citenamefont [1]{#1}%
\providecommand \href@noop [0]{\@secondoftwo}%
\providecommand \href [0]{\begingroup \@sanitize@url \@href}%
\providecommand \@href[1]{\@@startlink{#1}\@@href}%
\providecommand \@@href[1]{\endgroup#1\@@endlink}%
\providecommand \@sanitize@url [0]{\catcode `\\12\catcode `\$12\catcode
  `\&12\catcode `\#12\catcode `\^12\catcode `\_12\catcode `\%12\relax}%
\providecommand \@@startlink[1]{}%
\providecommand \@@endlink[0]{}%
\providecommand \url  [0]{\begingroup\@sanitize@url \@url }%
\providecommand \@url [1]{\endgroup\@href {#1}{\urlprefix }}%
\providecommand \urlprefix  [0]{URL }%
\providecommand \Eprint [0]{\href }%
\providecommand \doibase [0]{https://doi.org/}%
\providecommand \selectlanguage [0]{\@gobble}%
\providecommand \bibinfo  [0]{\@secondoftwo}%
\providecommand \bibfield  [0]{\@secondoftwo}%
\providecommand \translation [1]{[#1]}%
\providecommand \BibitemOpen [0]{}%
\providecommand \bibitemStop [0]{}%
\providecommand \bibitemNoStop [0]{.\EOS\space}%
\providecommand \EOS [0]{\spacefactor3000\relax}%
\providecommand \BibitemShut  [1]{\csname bibitem#1\endcsname}%
\let\auto@bib@innerbib\@empty
\bibitem [{\citenamefont {Allsman}\ \emph {et~al.}(2001)\citenamefont {Allsman}
  \emph {et~al.}}]{MACHO2001}%
  \BibitemOpen
  \bibfield  {author} {\bibinfo {author} {\bibfnamefont {R.}~\bibnamefont
  {Allsman}} \emph {et~al.} (\bibinfo {collaboration} {{MACHO}}),\ }\bibfield
  {title} {\bibinfo {title} {{MACHO project limits on black hole dark matter in
  the 1-30 solar mass range}},\ }\href {https://doi.org/10.1086/319636}
  {\bibfield  {journal} {\bibinfo  {journal} {Astrophys. J. Lett.}\ }\textbf
  {\bibinfo {volume} {550}},\ \bibinfo {pages} {L169} (\bibinfo {year}
  {2001})}\BibitemShut {NoStop}%
\bibitem [{\citenamefont {Tisserand}\ \emph {et~al.}(2007)\citenamefont
  {Tisserand} \emph {et~al.}}]{EROSMACHO2007}%
  \BibitemOpen
  \bibfield  {author} {\bibinfo {author} {\bibfnamefont {P.}~\bibnamefont
  {Tisserand}} \emph {et~al.} (\bibinfo {collaboration} {EROS-2}),\ }\bibfield
  {title} {\bibinfo {title} {{Limits on the Macho Content of the Galactic Halo
  from the EROS-2 Survey of the Magellanic Clouds}},\ }\href
  {https://doi.org/10.1051/0004-6361:20066017} {\bibfield  {journal} {\bibinfo
  {journal} {Astron. Astrophys.}\ }\textbf {\bibinfo {volume} {469}},\ \bibinfo
  {pages} {387} (\bibinfo {year} {2007})}\BibitemShut {NoStop}%
\bibitem [{\citenamefont {Wyrzykowski}\ \emph {et~al.}(2011)\citenamefont
  {Wyrzykowski} \emph {et~al.}}]{OGLE2011}%
  \BibitemOpen
  \bibfield  {author} {\bibinfo {author} {\bibfnamefont {L.}~\bibnamefont
  {Wyrzykowski}} \emph {et~al.},\ }\bibfield  {title} {\bibinfo {title} {{The
  OGLE View of Microlensing towards the Magellanic Clouds. IV. OGLE-III SMC
  Data and Final Conclusions on MACHOs}},\ }\href
  {https://doi.org/10.1111/j.1365-2966.2011.19243.x} {\bibfield  {journal}
  {\bibinfo  {journal} {Mon. Not. Roy. Astron. Soc.}\ }\textbf {\bibinfo
  {volume} {416}},\ \bibinfo {pages} {2949} (\bibinfo {year}
  {2011})}\BibitemShut {NoStop}%
\bibitem [{\citenamefont {Niikura}\ \emph {et~al.}(2019)\citenamefont {Niikura}
  \emph {et~al.}}]{HSCSubaru2017}%
  \BibitemOpen
  \bibfield  {author} {\bibinfo {author} {\bibfnamefont {H.}~\bibnamefont
  {Niikura}} \emph {et~al.},\ }\bibfield  {title} {\bibinfo {title}
  {Microlensing constraints on primordial black holes with {Subaru/HSC
  Andromeda} observations},\ }\href {https://doi.org/10.1038/s41550-019-0723-1}
  {\bibfield  {journal} {\bibinfo  {journal} {Nature Astronomy}\ }\textbf
  {\bibinfo {volume} {3}},\ \bibinfo {pages} {524} (\bibinfo {year}
  {2019})}\BibitemShut {NoStop}%
\bibitem [{\citenamefont {Carr}\ \emph
  {et~al.}(2010{\natexlab{a}})\citenamefont {Carr}, \citenamefont {Kohri},
  \citenamefont {Sendouda},\ and\ \citenamefont {Yokoyama}}]{PBHEvaporation}%
  \BibitemOpen
  \bibfield  {author} {\bibinfo {author} {\bibfnamefont {B.~J.}\ \bibnamefont
  {Carr}}, \bibinfo {author} {\bibfnamefont {K.}~\bibnamefont {Kohri}},
  \bibinfo {author} {\bibfnamefont {Y.}~\bibnamefont {Sendouda}},\ and\
  \bibinfo {author} {\bibfnamefont {J.}~\bibnamefont {Yokoyama}},\ }\bibfield
  {title} {\bibinfo {title} {{New Cosmological Constraints on Primordial Black
  Holes}},\ }\href {https://doi.org/10.1103/PhysRevD.81.104019} {\bibfield
  {journal} {\bibinfo  {journal} {Phys. Rev. D}\ }\textbf {\bibinfo {volume}
  {81}},\ \bibinfo {pages} {104019} (\bibinfo {year}
  {2010}{\natexlab{a}})}\BibitemShut {NoStop}%
\bibitem [{\citenamefont {Capela}\ \emph
  {et~al.}(2013{\natexlab{a}})\citenamefont {Capela}, \citenamefont
  {Pshirkov},\ and\ \citenamefont {Tinyakov}}]{PBHNeutronStarCapture}%
  \BibitemOpen
  \bibfield  {author} {\bibinfo {author} {\bibfnamefont {F.}~\bibnamefont
  {Capela}}, \bibinfo {author} {\bibfnamefont {M.}~\bibnamefont {Pshirkov}},\
  and\ \bibinfo {author} {\bibfnamefont {P.}~\bibnamefont {Tinyakov}},\
  }\bibfield  {title} {\bibinfo {title} {{Constraints on primordial black holes
  as dark matter candidates from capture by neutron stars}},\ }\href
  {https://doi.org/10.1103/PhysRevD.87.123524} {\bibfield  {journal} {\bibinfo
  {journal} {Phys.\ Rev.\ D}\ }\textbf {\bibinfo {volume} {87}},\ \bibinfo
  {pages} {123524} (\bibinfo {year} {2013}{\natexlab{a}})},\ \Eprint
  {https://arxiv.org/abs/1301.4984} {arXiv:1301.4984 [astro-ph.CO]}
  \BibitemShut {NoStop}%
\bibitem [{\citenamefont {Capela}\ \emph
  {et~al.}(2013{\natexlab{b}})\citenamefont {Capela}, \citenamefont
  {Pshirkov},\ and\ \citenamefont {Tinyakov}}]{PBHNS2013}%
  \BibitemOpen
  \bibfield  {author} {\bibinfo {author} {\bibfnamefont {F.}~\bibnamefont
  {Capela}}, \bibinfo {author} {\bibfnamefont {M.}~\bibnamefont {Pshirkov}},\
  and\ \bibinfo {author} {\bibfnamefont {P.}~\bibnamefont {Tinyakov}},\
  }\bibfield  {title} {\bibinfo {title} {Constraints on primordial black holes
  as dark matter candidates from star formation},\ }\href
  {https://doi.org/10.1103/PhysRevD.87.023507} {\bibfield  {journal} {\bibinfo
  {journal} {Phys. Rev. D}\ }\textbf {\bibinfo {volume} {87}},\ \bibinfo
  {pages} {023507} (\bibinfo {year} {2013}{\natexlab{b}})}\BibitemShut
  {NoStop}%
\bibitem [{\citenamefont {Capela}\ \emph
  {et~al.}(2013{\natexlab{c}})\citenamefont {Capela}, \citenamefont
  {Pshirkov},\ and\ \citenamefont {Tinyakov}}]{PBHNS20132}%
  \BibitemOpen
  \bibfield  {author} {\bibinfo {author} {\bibfnamefont {F.}~\bibnamefont
  {Capela}}, \bibinfo {author} {\bibfnamefont {M.}~\bibnamefont {Pshirkov}},\
  and\ \bibinfo {author} {\bibfnamefont {P.}~\bibnamefont {Tinyakov}},\
  }\bibfield  {title} {\bibinfo {title} {Constraints on primordial black holes
  as dark matter candidates from capture by neutron stars},\ }\href
  {https://doi.org/10.1103/PhysRevD.87.123524} {\bibfield  {journal} {\bibinfo
  {journal} {Phys. Rev. D}\ }\textbf {\bibinfo {volume} {87}},\ \bibinfo
  {pages} {123524} (\bibinfo {year} {2013}{\natexlab{c}})}\BibitemShut
  {NoStop}%
\bibitem [{\citenamefont {Graham}\ \emph {et~al.}(2015)\citenamefont {Graham},
  \citenamefont {Rajendran},\ and\ \citenamefont {Varela}}]{PBHWD2015}%
  \BibitemOpen
  \bibfield  {author} {\bibinfo {author} {\bibfnamefont {P.~W.}\ \bibnamefont
  {Graham}}, \bibinfo {author} {\bibfnamefont {S.}~\bibnamefont {Rajendran}},\
  and\ \bibinfo {author} {\bibfnamefont {J.}~\bibnamefont {Varela}},\
  }\bibfield  {title} {\bibinfo {title} {Dark matter triggers of supernovae},\
  }\href {https://doi.org/10.1103/PhysRevD.92.063007} {\bibfield  {journal}
  {\bibinfo  {journal} {Phys. Rev. D}\ }\textbf {\bibinfo {volume} {92}},\
  \bibinfo {pages} {063007} (\bibinfo {year} {2015})}\BibitemShut {NoStop}%
\bibitem [{\citenamefont {Boudaud}\ and\ \citenamefont
  {Cirelli}(2019)}]{PBHVoyager2018}%
  \BibitemOpen
  \bibfield  {author} {\bibinfo {author} {\bibfnamefont {M.}~\bibnamefont
  {Boudaud}}\ and\ \bibinfo {author} {\bibfnamefont {M.}~\bibnamefont
  {Cirelli}},\ }\bibfield  {title} {\bibinfo {title} {Voyager 1
  ${e}^{\ifmmode\pm\else\textpm\fi{}}$ further constrain primordial black holes
  as dark matter},\ }\href {https://doi.org/10.1103/PhysRevLett.122.041104}
  {\bibfield  {journal} {\bibinfo  {journal} {Phys. Rev. Lett.}\ }\textbf
  {\bibinfo {volume} {122}},\ \bibinfo {pages} {041104} (\bibinfo {year}
  {2019})}\BibitemShut {NoStop}%
\bibitem [{\citenamefont {DeRocco}\ and\ \citenamefont
  {Graham}(2019)}]{PBHWill}%
  \BibitemOpen
  \bibfield  {author} {\bibinfo {author} {\bibfnamefont {W.}~\bibnamefont
  {DeRocco}}\ and\ \bibinfo {author} {\bibfnamefont {P.~W.}\ \bibnamefont
  {Graham}},\ }\bibfield  {title} {\bibinfo {title} {Constraining primordial
  black hole abundance with the galactic 511 kev line},\ }\href
  {https://doi.org/10.1103/PhysRevLett.123.251102} {\bibfield  {journal}
  {\bibinfo  {journal} {Phys. Rev. Lett.}\ }\textbf {\bibinfo {volume} {123}},\
  \bibinfo {pages} {251102} (\bibinfo {year} {2019})}\BibitemShut {NoStop}%
\bibitem [{\citenamefont {Katz}\ \emph {et~al.}(2018)\citenamefont {Katz},
  \citenamefont {Kopp}, \citenamefont {Sibiryakov},\ and\ \citenamefont
  {Xue}}]{Katz_2018}%
  \BibitemOpen
  \bibfield  {author} {\bibinfo {author} {\bibfnamefont {A.}~\bibnamefont
  {Katz}}, \bibinfo {author} {\bibfnamefont {J.}~\bibnamefont {Kopp}}, \bibinfo
  {author} {\bibfnamefont {S.}~\bibnamefont {Sibiryakov}},\ and\ \bibinfo
  {author} {\bibfnamefont {W.}~\bibnamefont {Xue}},\ }\bibfield  {title}
  {\bibinfo {title} {Femtolensing by dark matter revisited},\ }\href
  {https://doi.org/10.1088/1475-7516/2018/12/005} {\bibfield  {journal}
  {\bibinfo  {journal} {Journal of Cosmology and Astroparticle Physics}\
  }\textbf {\bibinfo {volume} {2018}}\bibinfo  {number} { (12)},\ \bibinfo
  {pages} {005}}\BibitemShut {NoStop}%
\bibitem [{\citenamefont {Peccei}\ and\ \citenamefont
  {Quinn}(1977{\natexlab{a}})}]{PQ1}%
  \BibitemOpen
\bibfield  {number} {  }\bibfield  {author} {\bibinfo {author} {\bibfnamefont
  {R.~D.}\ \bibnamefont {Peccei}}\ and\ \bibinfo {author} {\bibfnamefont
  {H.~R.}\ \bibnamefont {Quinn}},\ }\bibfield  {title} {\bibinfo {title}
  {$\mathrm{CP}$ conservation in the presence of pseudoparticles},\ }\href
  {https://doi.org/10.1103/PhysRevLett.38.1440} {\bibfield  {journal} {\bibinfo
   {journal} {Phys. Rev. Lett.}\ }\textbf {\bibinfo {volume} {38}},\ \bibinfo
  {pages} {1440} (\bibinfo {year} {1977}{\natexlab{a}})}\BibitemShut {NoStop}%
\bibitem [{\citenamefont {Peccei}\ and\ \citenamefont
  {Quinn}(1977{\natexlab{b}})}]{PQ2}%
  \BibitemOpen
  \bibfield  {author} {\bibinfo {author} {\bibfnamefont {R.~D.}\ \bibnamefont
  {Peccei}}\ and\ \bibinfo {author} {\bibfnamefont {H.~R.}\ \bibnamefont
  {Quinn}},\ }\bibfield  {title} {\bibinfo {title} {Constraints imposed by
  $\mathrm{CP}$ conservation in the presence of pseudoparticles},\ }\href
  {https://doi.org/10.1103/PhysRevD.16.1791} {\bibfield  {journal} {\bibinfo
  {journal} {Phys. Rev. D}\ }\textbf {\bibinfo {volume} {16}},\ \bibinfo
  {pages} {1791} (\bibinfo {year} {1977}{\natexlab{b}})}\BibitemShut {NoStop}%
\bibitem [{\citenamefont {Weinberg}(1978)}]{WeinbergAxion}%
  \BibitemOpen
  \bibfield  {author} {\bibinfo {author} {\bibfnamefont {S.}~\bibnamefont
  {Weinberg}},\ }\bibfield  {title} {\bibinfo {title} {A new light boson?},\
  }\href {https://doi.org/10.1103/PhysRevLett.40.223} {\bibfield  {journal}
  {\bibinfo  {journal} {Phys. Rev. Lett.}\ }\textbf {\bibinfo {volume} {40}},\
  \bibinfo {pages} {223} (\bibinfo {year} {1978})}\BibitemShut {NoStop}%
\bibitem [{\citenamefont {Wilczek}(1978)}]{WilczekAxion}%
  \BibitemOpen
  \bibfield  {author} {\bibinfo {author} {\bibfnamefont {F.}~\bibnamefont
  {Wilczek}},\ }\bibfield  {title} {\bibinfo {title} {Problem of strong {$P$}
  and {$T$} invariance in the presence of instantons},\ }\href
  {https://doi.org/10.1103/PhysRevLett.40.279} {\bibfield  {journal} {\bibinfo
  {journal} {Phys. Rev. Lett.}\ }\textbf {\bibinfo {volume} {40}},\ \bibinfo
  {pages} {279} (\bibinfo {year} {1978})}\BibitemShut {NoStop}%
\bibitem [{\citenamefont {Abbott}\ and\ \citenamefont
  {Sikivie}(1983)}]{Abbott1982}%
  \BibitemOpen
  \bibfield  {author} {\bibinfo {author} {\bibfnamefont {L.}~\bibnamefont
  {Abbott}}\ and\ \bibinfo {author} {\bibfnamefont {P.}~\bibnamefont
  {Sikivie}},\ }\bibfield  {title} {\bibinfo {title} {{A Cosmological Bound on
  the Invisible Axion}},\ }\href {https://doi.org/10.1016/0370-2693(83)90638-X}
  {\bibfield  {journal} {\bibinfo  {journal} {Phys. Lett. B}\ }\textbf
  {\bibinfo {volume} {120}},\ \bibinfo {pages} {133} (\bibinfo {year}
  {1983})}\BibitemShut {NoStop}%
\bibitem [{\citenamefont {Dine}\ and\ \citenamefont
  {Fischler}(1983)}]{Fischler1982}%
  \BibitemOpen
  \bibfield  {author} {\bibinfo {author} {\bibfnamefont {M.}~\bibnamefont
  {Dine}}\ and\ \bibinfo {author} {\bibfnamefont {W.}~\bibnamefont
  {Fischler}},\ }\bibfield  {title} {\bibinfo {title} {{The Not So Harmless
  Axion}},\ }\href {https://doi.org/10.1016/0370-2693(83)90639-1} {\bibfield
  {journal} {\bibinfo  {journal} {Phys. Lett. B}\ }\textbf {\bibinfo {volume}
  {120}},\ \bibinfo {pages} {137} (\bibinfo {year} {1983})}\BibitemShut
  {NoStop}%
\bibitem [{\citenamefont {Preskill}\ \emph {et~al.}(1983)\citenamefont
  {Preskill}, \citenamefont {Wise},\ and\ \citenamefont
  {Wilczek}}]{PRESKILL1983127}%
  \BibitemOpen
  \bibfield  {author} {\bibinfo {author} {\bibfnamefont {J.}~\bibnamefont
  {Preskill}}, \bibinfo {author} {\bibfnamefont {M.~B.}\ \bibnamefont {Wise}},\
  and\ \bibinfo {author} {\bibfnamefont {F.}~\bibnamefont {Wilczek}},\
  }\bibfield  {title} {\bibinfo {title} {Cosmology of the invisible axion},\
  }\href {https://doi.org/https://doi.org/10.1016/0370-2693(83)90637-8}
  {\bibfield  {journal} {\bibinfo  {journal} {Physics Letters B}\ }\textbf
  {\bibinfo {volume} {120}},\ \bibinfo {pages} {127 } (\bibinfo {year}
  {1983})}\BibitemShut {NoStop}%
\bibitem [{\citenamefont {Arvanitaki}\ \emph {et~al.}(2010)\citenamefont
  {Arvanitaki}, \citenamefont {Dimopoulos}, \citenamefont {Dubovsky},
  \citenamefont {Kaloper},\ and\ \citenamefont {March-Russell}}]{Axiverse2010}%
  \BibitemOpen
  \bibfield  {author} {\bibinfo {author} {\bibfnamefont {A.}~\bibnamefont
  {Arvanitaki}}, \bibinfo {author} {\bibfnamefont {S.}~\bibnamefont
  {Dimopoulos}}, \bibinfo {author} {\bibfnamefont {S.}~\bibnamefont
  {Dubovsky}}, \bibinfo {author} {\bibfnamefont {N.}~\bibnamefont {Kaloper}},\
  and\ \bibinfo {author} {\bibfnamefont {J.}~\bibnamefont {March-Russell}},\
  }\bibfield  {title} {\bibinfo {title} {String axiverse},\ }\href
  {https://doi.org/10.1103/PhysRevD.81.123530} {\bibfield  {journal} {\bibinfo
  {journal} {Phys. Rev. D}\ }\textbf {\bibinfo {volume} {81}},\ \bibinfo
  {pages} {123530} (\bibinfo {year} {2010})}\BibitemShut {NoStop}%
\bibitem [{\citenamefont {Arvanitaki}\ \emph {et~al.}(2020)\citenamefont
  {Arvanitaki}, \citenamefont {Dimopoulos}, \citenamefont {Galanis},
  \citenamefont {Lehner}, \citenamefont {Thompson},\ and\ \citenamefont
  {Van~Tilburg}}]{marios2019}%
  \BibitemOpen
  \bibfield  {author} {\bibinfo {author} {\bibfnamefont {A.}~\bibnamefont
  {Arvanitaki}}, \bibinfo {author} {\bibfnamefont {S.}~\bibnamefont
  {Dimopoulos}}, \bibinfo {author} {\bibfnamefont {M.}~\bibnamefont {Galanis}},
  \bibinfo {author} {\bibfnamefont {L.}~\bibnamefont {Lehner}}, \bibinfo
  {author} {\bibfnamefont {J.~O.}\ \bibnamefont {Thompson}},\ and\ \bibinfo
  {author} {\bibfnamefont {K.}~\bibnamefont {Van~Tilburg}},\ }\bibfield
  {title} {\bibinfo {title} {Large-misalignment mechanism for the formation of
  compact axion structures: Signatures from the {QCD} axion to fuzzy dark
  matter},\ }\href {https://doi.org/10.1103/PhysRevD.101.083014} {\bibfield
  {journal} {\bibinfo  {journal} {Phys. Rev. D}\ }\textbf {\bibinfo {volume}
  {101}},\ \bibinfo {pages} {083014} (\bibinfo {year} {2020})}\BibitemShut
  {NoStop}%
\bibitem [{\citenamefont {Iwazaki}(2014)}]{iwazaki2014}%
  \BibitemOpen
  \bibfield  {author} {\bibinfo {author} {\bibfnamefont {A.}~\bibnamefont
  {Iwazaki}},\ }\href@noop {} {\bibinfo {title} {Fast radio bursts from axion
  stars}} (\bibinfo {year} {2014}),\ \Eprint {https://arxiv.org/abs/1412.7825}
  {arXiv:1412.7825 [hep-ph]} \BibitemShut {NoStop}%
\bibitem [{\citenamefont {Iwazaki}(2015)}]{Iwazaki2015}%
  \BibitemOpen
  \bibfield  {author} {\bibinfo {author} {\bibfnamefont {A.}~\bibnamefont
  {Iwazaki}},\ }\bibfield  {title} {\bibinfo {title} {Axion stars and fast
  radio bursts},\ }\href {https://doi.org/10.1103/PhysRevD.91.023008}
  {\bibfield  {journal} {\bibinfo  {journal} {Phys. Rev. D}\ }\textbf {\bibinfo
  {volume} {91}},\ \bibinfo {pages} {023008} (\bibinfo {year}
  {2015})}\BibitemShut {NoStop}%
\bibitem [{\citenamefont {Raby}(2016)}]{Raby2016}%
  \BibitemOpen
  \bibfield  {author} {\bibinfo {author} {\bibfnamefont {S.}~\bibnamefont
  {Raby}},\ }\bibfield  {title} {\bibinfo {title} {Axion star collisions with
  neutron stars and fast radio bursts},\ }\href
  {https://doi.org/10.1103/PhysRevD.94.103004} {\bibfield  {journal} {\bibinfo
  {journal} {Phys. Rev. D}\ }\textbf {\bibinfo {volume} {94}},\ \bibinfo
  {pages} {103004} (\bibinfo {year} {2016})}\BibitemShut {NoStop}%
\bibitem [{\citenamefont {Bai}\ and\ \citenamefont {Hamada}(2018)}]{bai2018}%
  \BibitemOpen
  \bibfield  {author} {\bibinfo {author} {\bibfnamefont {Y.}~\bibnamefont
  {Bai}}\ and\ \bibinfo {author} {\bibfnamefont {Y.}~\bibnamefont {Hamada}},\
  }\bibfield  {title} {\bibinfo {title} {Detecting axion stars with radio
  telescopes},\ }\href
  {https://doi.org/https://doi.org/10.1016/j.physletb.2018.03.070} {\bibfield
  {journal} {\bibinfo  {journal} {Physics Letters B}\ }\textbf {\bibinfo
  {volume} {781}},\ \bibinfo {pages} {187 } (\bibinfo {year}
  {2018})}\BibitemShut {NoStop}%
\bibitem [{\citenamefont {Prabhu}\ and\ \citenamefont
  {Rapidis}(2020)}]{prabhu2020resonant}%
  \BibitemOpen
  \bibfield  {author} {\bibinfo {author} {\bibfnamefont {A.}~\bibnamefont
  {Prabhu}}\ and\ \bibinfo {author} {\bibfnamefont {N.}~\bibnamefont
  {Rapidis}},\ }\href@noop {} {\bibinfo {title} {Resonant conversion of dark
  matter oscillons in pulsar magnetospheres}} (\bibinfo {year} {2020}),\
  \Eprint {https://arxiv.org/abs/2005.03700} {arXiv:2005.03700 [astro-ph.CO]}
  \BibitemShut {NoStop}%
\bibitem [{\citenamefont {Fairbairn}\ \emph {et~al.}(2018)\citenamefont
  {Fairbairn}, \citenamefont {Marsh}, \citenamefont {Quevillon},\ and\
  \citenamefont {Rozier}}]{fairbairn2018microlensing}%
  \BibitemOpen
  \bibfield  {author} {\bibinfo {author} {\bibfnamefont {M.}~\bibnamefont
  {Fairbairn}}, \bibinfo {author} {\bibfnamefont {D.~J.~E.}\ \bibnamefont
  {Marsh}}, \bibinfo {author} {\bibfnamefont {J.}~\bibnamefont {Quevillon}},\
  and\ \bibinfo {author} {\bibfnamefont {S.}~\bibnamefont {Rozier}},\
  }\bibfield  {title} {\bibinfo {title} {Structure formation and microlensing
  with axion miniclusters},\ }\href
  {https://doi.org/10.1103/PhysRevD.97.083502} {\bibfield  {journal} {\bibinfo
  {journal} {Phys. Rev. D}\ }\textbf {\bibinfo {volume} {97}},\ \bibinfo
  {pages} {083502} (\bibinfo {year} {2018})}\BibitemShut {NoStop}%
\bibitem [{\citenamefont {Kolb}\ and\ \citenamefont
  {Tkachev}(1996)}]{KolbTkachevPicoFemtolensing}%
  \BibitemOpen
  \bibfield  {author} {\bibinfo {author} {\bibfnamefont {E.~W.}\ \bibnamefont
  {Kolb}}\ and\ \bibinfo {author} {\bibfnamefont {I.~I.}\ \bibnamefont
  {Tkachev}},\ }\bibfield  {title} {\bibinfo {title} {Femtolensing and
  picolensing by axion miniclusters},\ }\bibfield  {journal} {\bibinfo
  {journal} {The Astrophysical Journal}\ }\textbf {\bibinfo {volume} {460}},\
  \href {https://doi.org/10.1086/309962} {10.1086/309962} (\bibinfo {year}
  {1996})\BibitemShut {NoStop}%
\bibitem [{\citenamefont {Fairbairn}\ \emph {et~al.}(2017)\citenamefont
  {Fairbairn}, \citenamefont {Marsh},\ and\ \citenamefont
  {Quevillon}}]{Fairbairn2017}%
  \BibitemOpen
  \bibfield  {author} {\bibinfo {author} {\bibfnamefont {M.}~\bibnamefont
  {Fairbairn}}, \bibinfo {author} {\bibfnamefont {D.~J.~E.}\ \bibnamefont
  {Marsh}},\ and\ \bibinfo {author} {\bibfnamefont {J.}~\bibnamefont
  {Quevillon}},\ }\bibfield  {title} {\bibinfo {title} {Searching for the {QCD}
  axion with gravitational microlensing},\ }\href
  {https://doi.org/10.1103/PhysRevLett.119.021101} {\bibfield  {journal}
  {\bibinfo  {journal} {Phys. Rev. Lett.}\ }\textbf {\bibinfo {volume} {119}},\
  \bibinfo {pages} {021101} (\bibinfo {year} {2017})}\BibitemShut {NoStop}%
\bibitem [{\citenamefont {McDonald}\ and\ \citenamefont
  {Ventura}(2020)}]{mcdonald2019optical}%
  \BibitemOpen
  \bibfield  {author} {\bibinfo {author} {\bibfnamefont {J.~I.}\ \bibnamefont
  {McDonald}}\ and\ \bibinfo {author} {\bibfnamefont {L.~B.}\ \bibnamefont
  {Ventura}},\ }\bibfield  {title} {\bibinfo {title} {Optical properties of
  dynamical axion backgrounds},\ }\href
  {https://doi.org/10.1103/PhysRevD.101.123503} {\bibfield  {journal} {\bibinfo
   {journal} {Phys. Rev. D}\ }\textbf {\bibinfo {volume} {101}},\ \bibinfo
  {pages} {123503} (\bibinfo {year} {2020})}\BibitemShut {NoStop}%
\bibitem [{\citenamefont {Mcdonald}\ and\ \citenamefont
  {Ventura}(2020)}]{McDonald2020}%
  \BibitemOpen
  \bibfield  {author} {\bibinfo {author} {\bibfnamefont {J.~I.}\ \bibnamefont
  {Mcdonald}}\ and\ \bibinfo {author} {\bibfnamefont {L.~B.}\ \bibnamefont
  {Ventura}},\ }\bibfield  {title} {\bibinfo {title} {{Bending of light in
  axion backgrounds}},\ }\href@noop {} {\  (\bibinfo {year} {2020})},\ \Eprint
  {https://arxiv.org/abs/2008.12923} {arXiv:2008.12923 [hep-ph]} \BibitemShut
  {NoStop}%
\bibitem [{\citenamefont {Farina}\ \emph {et~al.}(2017)\citenamefont {Farina},
  \citenamefont {Pappadopulo}, \citenamefont {Rompineve},\ and\ \citenamefont
  {Tesi}}]{photophilic2016}%
  \BibitemOpen
  \bibfield  {author} {\bibinfo {author} {\bibfnamefont {M.}~\bibnamefont
  {Farina}}, \bibinfo {author} {\bibfnamefont {D.}~\bibnamefont {Pappadopulo}},
  \bibinfo {author} {\bibfnamefont {F.}~\bibnamefont {Rompineve}},\ and\
  \bibinfo {author} {\bibfnamefont {A.}~\bibnamefont {Tesi}},\ }\bibfield
  {title} {\bibinfo {title} {The photo-philic {QCD} axion},\ }\href
  {https://doi.org/10.1007/JHEP01(2017)095} {\bibfield  {journal} {\bibinfo
  {journal} {Journal of High Energy Physics}\ }\textbf {\bibinfo {volume}
  {2017}},\ \bibinfo {pages} {95} (\bibinfo {year} {2017})}\BibitemShut
  {NoStop}%
\bibitem [{\citenamefont {Carroll}\ \emph {et~al.}(1990)\citenamefont
  {Carroll}, \citenamefont {Field},\ and\ \citenamefont {Jackiw}}]{jackiw}%
  \BibitemOpen
  \bibfield  {author} {\bibinfo {author} {\bibfnamefont {S.~M.}\ \bibnamefont
  {Carroll}}, \bibinfo {author} {\bibfnamefont {G.~B.}\ \bibnamefont {Field}},\
  and\ \bibinfo {author} {\bibfnamefont {R.}~\bibnamefont {Jackiw}},\
  }\bibfield  {title} {\bibinfo {title} {Limits on a lorentz- and
  parity-violating modification of electrodynamics},\ }\href
  {https://doi.org/10.1103/PhysRevD.41.1231} {\bibfield  {journal} {\bibinfo
  {journal} {Phys. Rev. D}\ }\textbf {\bibinfo {volume} {41}},\ \bibinfo
  {pages} {1231} (\bibinfo {year} {1990})}\BibitemShut {NoStop}%
\bibitem [{\citenamefont {Dong}\ \emph {et~al.}(2011)\citenamefont {Dong},
  \citenamefont {Horn}, \citenamefont {Silverstein},\ and\ \citenamefont
  {Westphal}}]{Eva2011}%
  \BibitemOpen
  \bibfield  {author} {\bibinfo {author} {\bibfnamefont {X.}~\bibnamefont
  {Dong}}, \bibinfo {author} {\bibfnamefont {B.}~\bibnamefont {Horn}}, \bibinfo
  {author} {\bibfnamefont {E.}~\bibnamefont {Silverstein}},\ and\ \bibinfo
  {author} {\bibfnamefont {A.}~\bibnamefont {Westphal}},\ }\bibfield  {title}
  {\bibinfo {title} {Simple exercises to flatten your potential},\ }\href
  {https://doi.org/10.1103/PhysRevD.84.026011} {\bibfield  {journal} {\bibinfo
  {journal} {Phys. Rev. D}\ }\textbf {\bibinfo {volume} {84}},\ \bibinfo
  {pages} {026011} (\bibinfo {year} {2011})}\BibitemShut {NoStop}%
\bibitem [{\citenamefont {Brillouin}(1960)}]{brillouin}%
  \BibitemOpen
  \bibfield  {author} {\bibinfo {author} {\bibfnamefont {L.}~\bibnamefont
  {Brillouin}},\ }\href
  {https://doi.org/https://doi.org/10.1016/B978-1-4832-3068-9.50001-5} {\emph
  {\bibinfo {title} {Wave Propagation and Group Velocity}}},\ \bibinfo {series}
  {Pure and Applied Physics}, Vol.~\bibinfo {volume} {8}\ (\bibinfo
  {publisher} {Elsevier},\ \bibinfo {year} {1960})\BibitemShut {NoStop}%
\bibitem [{\citenamefont {Garrison}\ \emph {et~al.}(1998)\citenamefont
  {Garrison}, \citenamefont {Mitchell}, \citenamefont {Chiao},\ and\
  \citenamefont {Bolda}}]{SuperluminalDebate}%
  \BibitemOpen
  \bibfield  {author} {\bibinfo {author} {\bibfnamefont {J.}~\bibnamefont
  {Garrison}}, \bibinfo {author} {\bibfnamefont {M.}~\bibnamefont {Mitchell}},
  \bibinfo {author} {\bibfnamefont {R.}~\bibnamefont {Chiao}},\ and\ \bibinfo
  {author} {\bibfnamefont {E.}~\bibnamefont {Bolda}},\ }\bibfield  {title}
  {\bibinfo {title} {{Superluminal signals: Causal loop paradoxes revisited}},\
  }\href {https://doi.org/10.1016/S0375-9601(98)00381-8} {\bibfield  {journal}
  {\bibinfo  {journal} {Phys. Lett. A}\ }\textbf {\bibinfo {volume} {245}},\
  \bibinfo {pages} {19} (\bibinfo {year} {1998})},\ \Eprint
  {https://arxiv.org/abs/quant-ph/9810031} {arXiv:quant-ph/9810031}
  \BibitemShut {NoStop}%
\bibitem [{\citenamefont {Milonni}(2005)}]{MilonniSuperluminal}%
  \BibitemOpen
  \bibfield  {author} {\bibinfo {author} {\bibfnamefont {P.~W.}\ \bibnamefont
  {Milonni}},\ }\href@noop {} {\emph {\bibinfo {title} {Fast Light, Slow Light
  and Left-Handed Light}}}\ (\bibinfo  {publisher} {Institute of Physics
  Publishing},\ \bibinfo {year} {2005})\BibitemShut {NoStop}%
\bibitem [{\citenamefont {Weinberg}(1962)}]{WeinbergEikonal1962}%
  \BibitemOpen
  \bibfield  {author} {\bibinfo {author} {\bibfnamefont {S.}~\bibnamefont
  {Weinberg}},\ }\bibfield  {title} {\bibinfo {title} {Eikonal method in
  magnetohydrodynamics},\ }\href {https://doi.org/10.1103/PhysRev.126.1899}
  {\bibfield  {journal} {\bibinfo  {journal} {Phys. Rev.}\ }\textbf {\bibinfo
  {volume} {126}},\ \bibinfo {pages} {1899} (\bibinfo {year}
  {1962})}\BibitemShut {NoStop}%
\bibitem [{\citenamefont {Plascencia}\ and\ \citenamefont
  {Urbano}(2018)}]{Plascencia_2018}%
  \BibitemOpen
  \bibfield  {author} {\bibinfo {author} {\bibfnamefont {A.~D.}\ \bibnamefont
  {Plascencia}}\ and\ \bibinfo {author} {\bibfnamefont {A.}~\bibnamefont
  {Urbano}},\ }\bibfield  {title} {\bibinfo {title} {Black hole superradiance
  and polarization-dependent bending of light},\ }\href
  {https://doi.org/10.1088/1475-7516/2018/04/059} {\bibfield  {journal}
  {\bibinfo  {journal} {Journal of Cosmology and Astroparticle Physics}\
  }\textbf {\bibinfo {volume} {2018}}\bibinfo  {number} { (04)},\ \bibinfo
  {pages} {059}}\BibitemShut {NoStop}%
\bibitem [{\citenamefont {Blas}\ \emph {et~al.}(2020)\citenamefont {Blas},
  \citenamefont {Caputo}, \citenamefont {Ivanov},\ and\ \citenamefont
  {Sberna}}]{blas2019chiral}%
  \BibitemOpen
\bibfield  {number} {  }\bibfield  {author} {\bibinfo {author} {\bibfnamefont
  {D.}~\bibnamefont {Blas}}, \bibinfo {author} {\bibfnamefont {A.}~\bibnamefont
  {Caputo}}, \bibinfo {author} {\bibfnamefont {M.~M.}\ \bibnamefont {Ivanov}},\
  and\ \bibinfo {author} {\bibfnamefont {L.}~\bibnamefont {Sberna}},\
  }\bibfield  {title} {\bibinfo {title} {No chiral light bending by clumps of
  axion-like particles},\ }\href
  {https://doi.org/https://doi.org/10.1016/j.dark.2019.100428} {\bibfield
  {journal} {\bibinfo  {journal} {Physics of the Dark Universe}\ }\textbf
  {\bibinfo {volume} {27}},\ \bibinfo {pages} {100428} (\bibinfo {year}
  {2020})}\BibitemShut {NoStop}%
\bibitem [{\citenamefont {Levkov}\ \emph {et~al.}(2018)\citenamefont {Levkov},
  \citenamefont {Panin},\ and\ \citenamefont {Tkachev}}]{tkachev2018}%
  \BibitemOpen
  \bibfield  {author} {\bibinfo {author} {\bibfnamefont {D.~G.}\ \bibnamefont
  {Levkov}}, \bibinfo {author} {\bibfnamefont {A.~G.}\ \bibnamefont {Panin}},\
  and\ \bibinfo {author} {\bibfnamefont {I.~I.}\ \bibnamefont {Tkachev}},\
  }\bibfield  {title} {\bibinfo {title} {Gravitational bose-einstein
  condensation in the kinetic regime},\ }\href
  {https://doi.org/10.1103/PhysRevLett.121.151301} {\bibfield  {journal}
  {\bibinfo  {journal} {Phys. Rev. Lett.}\ }\textbf {\bibinfo {volume} {121}},\
  \bibinfo {pages} {151301} (\bibinfo {year} {2018})}\BibitemShut {NoStop}%
\bibitem [{\citenamefont {Gross}\ \emph {et~al.}(1981)\citenamefont {Gross},
  \citenamefont {Pisarski},\ and\ \citenamefont {Yaffe}}]{Gross1981}%
  \BibitemOpen
  \bibfield  {author} {\bibinfo {author} {\bibfnamefont {D.~J.}\ \bibnamefont
  {Gross}}, \bibinfo {author} {\bibfnamefont {R.~D.}\ \bibnamefont
  {Pisarski}},\ and\ \bibinfo {author} {\bibfnamefont {L.~G.}\ \bibnamefont
  {Yaffe}},\ }\bibfield  {title} {\bibinfo {title} {{QCD} and instantons at
  finite temperature},\ }\href {https://doi.org/10.1103/RevModPhys.53.43}
  {\bibfield  {journal} {\bibinfo  {journal} {Rev. Mod. Phys.}\ }\textbf
  {\bibinfo {volume} {53}},\ \bibinfo {pages} {43} (\bibinfo {year}
  {1981})}\BibitemShut {NoStop}%
\bibitem [{\citenamefont {Hogan}\ and\ \citenamefont {Rees}(1988)}]{Hogan1988}%
  \BibitemOpen
  \bibfield  {author} {\bibinfo {author} {\bibfnamefont {C.}~\bibnamefont
  {Hogan}}\ and\ \bibinfo {author} {\bibfnamefont {M.}~\bibnamefont {Rees}},\
  }\bibfield  {title} {\bibinfo {title} {Axion miniclusters},\ }\href
  {https://doi.org/https://doi.org/10.1016/0370-2693(88)91655-3} {\bibfield
  {journal} {\bibinfo  {journal} {Physics Letters B}\ }\textbf {\bibinfo
  {volume} {205}},\ \bibinfo {pages} {228 } (\bibinfo {year}
  {1988})}\BibitemShut {NoStop}%
\bibitem [{\citenamefont {Kolb}\ and\ \citenamefont
  {Tkachev}(1993)}]{KolbTkachev93}%
  \BibitemOpen
  \bibfield  {author} {\bibinfo {author} {\bibfnamefont {E.~W.}\ \bibnamefont
  {Kolb}}\ and\ \bibinfo {author} {\bibfnamefont {I.~I.}\ \bibnamefont
  {Tkachev}},\ }\bibfield  {title} {\bibinfo {title} {Axion miniclusters and
  bose stars},\ }\href {https://doi.org/10.1103/PhysRevLett.71.3051} {\bibfield
   {journal} {\bibinfo  {journal} {Phys. Rev. Lett.}\ }\textbf {\bibinfo
  {volume} {71}},\ \bibinfo {pages} {3051} (\bibinfo {year}
  {1993})}\BibitemShut {NoStop}%
\bibitem [{\citenamefont {Kolb}\ and\ \citenamefont
  {Tkachev}(1994{\natexlab{a}})}]{KolbTkachev941}%
  \BibitemOpen
  \bibfield  {author} {\bibinfo {author} {\bibfnamefont {E.~W.}\ \bibnamefont
  {Kolb}}\ and\ \bibinfo {author} {\bibfnamefont {I.~I.}\ \bibnamefont
  {Tkachev}},\ }\bibfield  {title} {\bibinfo {title} {Nonlinear axion dynamics
  and the formation of cosmological pseudosolitons},\ }\href
  {https://doi.org/10.1103/PhysRevD.49.5040} {\bibfield  {journal} {\bibinfo
  {journal} {Phys. Rev. D}\ }\textbf {\bibinfo {volume} {49}},\ \bibinfo
  {pages} {5040} (\bibinfo {year} {1994}{\natexlab{a}})}\BibitemShut {NoStop}%
\bibitem [{\citenamefont {Kolb}\ and\ \citenamefont
  {Tkachev}(1994{\natexlab{b}})}]{KolbTkachev942}%
  \BibitemOpen
  \bibfield  {author} {\bibinfo {author} {\bibfnamefont {E.~W.}\ \bibnamefont
  {Kolb}}\ and\ \bibinfo {author} {\bibfnamefont {I.~I.}\ \bibnamefont
  {Tkachev}},\ }\bibfield  {title} {\bibinfo {title} {Large-amplitude
  isothermal fluctuations and high-density dark-matter clumps},\ }\href
  {https://doi.org/10.1103/PhysRevD.50.769} {\bibfield  {journal} {\bibinfo
  {journal} {Phys. Rev. D}\ }\textbf {\bibinfo {volume} {50}},\ \bibinfo
  {pages} {769} (\bibinfo {year} {1994}{\natexlab{b}})}\BibitemShut {NoStop}%
\bibitem [{\citenamefont {Hardy}(2017)}]{edhardy}%
  \BibitemOpen
  \bibfield  {author} {\bibinfo {author} {\bibfnamefont {E.}~\bibnamefont
  {Hardy}},\ }\bibfield  {title} {\bibinfo {title} {{Miniclusters in the
  Axiverse}},\ }\href {https://doi.org/10.1007/JHEP02(2017)046} {\bibfield
  {journal} {\bibinfo  {journal} {JHEP}\ }\textbf {\bibinfo {volume} {02}},\
  \bibinfo {pages} {046}},\ \Eprint {https://arxiv.org/abs/1609.00208}
  {arXiv:1609.00208 [hep-ph]} \BibitemShut {NoStop}%
\bibitem [{\citenamefont {Co}\ \emph {et~al.}(2019)\citenamefont {Co},
  \citenamefont {Gonzalez},\ and\ \citenamefont {Harigaya}}]{raymondco}%
  \BibitemOpen
  \bibfield  {author} {\bibinfo {author} {\bibfnamefont {R.~T.}\ \bibnamefont
  {Co}}, \bibinfo {author} {\bibfnamefont {E.}~\bibnamefont {Gonzalez}},\ and\
  \bibinfo {author} {\bibfnamefont {K.}~\bibnamefont {Harigaya}},\ }\bibfield
  {title} {\bibinfo {title} {Axion misalignment driven to the hilltop},\ }\href
  {https://doi.org/10.1007/JHEP05(2019)163} {\bibfield  {journal} {\bibinfo
  {journal} {Journal of High Energy Physics}\ }\textbf {\bibinfo {volume}
  {2019}},\ \bibinfo {pages} {163} (\bibinfo {year} {2019})}\BibitemShut
  {NoStop}%
\bibitem [{\citenamefont {Takahashi}\ and\ \citenamefont
  {Yin}(2019)}]{yinwen2019}%
  \BibitemOpen
  \bibfield  {author} {\bibinfo {author} {\bibfnamefont {F.}~\bibnamefont
  {Takahashi}}\ and\ \bibinfo {author} {\bibfnamefont {W.}~\bibnamefont
  {Yin}},\ }\bibfield  {title} {\bibinfo {title} {{QCD axion on hilltop by a
  phase shift of $\pi$}},\ }\href {https://doi.org/10.1007/JHEP10(2019)120}
  {\bibfield  {journal} {\bibinfo  {journal} {JHEP}\ }\textbf {\bibinfo
  {volume} {10}},\ \bibinfo {pages} {120}},\ \Eprint
  {https://arxiv.org/abs/1908.06071} {arXiv:1908.06071 [hep-ph]} \BibitemShut
  {NoStop}%
\bibitem [{\citenamefont {Khlebnikov}\ and\ \citenamefont
  {Tkachev}(2000)}]{Khlebnikov1}%
  \BibitemOpen
  \bibfield  {author} {\bibinfo {author} {\bibfnamefont {S.}~\bibnamefont
  {Khlebnikov}}\ and\ \bibinfo {author} {\bibfnamefont {I.}~\bibnamefont
  {Tkachev}},\ }\bibfield  {title} {\bibinfo {title} {Quantum dew: Formation of
  quantum liquid in a nonequilibrium {Bose} gas},\ }\href
  {https://doi.org/10.1103/PhysRevD.61.083517} {\bibfield  {journal} {\bibinfo
  {journal} {Phys. Rev. D}\ }\textbf {\bibinfo {volume} {61}},\ \bibinfo
  {pages} {083517} (\bibinfo {year} {2000})}\BibitemShut {NoStop}%
\bibitem [{\citenamefont {Khlebnikov}(2000)}]{Khlebnikov2}%
  \BibitemOpen
  \bibfield  {author} {\bibinfo {author} {\bibfnamefont {S.}~\bibnamefont
  {Khlebnikov}},\ }\bibfield  {title} {\bibinfo {title} {Short-scale
  gravitational instability in a disordered bose gas},\ }\href
  {https://doi.org/10.1103/PhysRevD.62.043519} {\bibfield  {journal} {\bibinfo
  {journal} {Phys. Rev. D}\ }\textbf {\bibinfo {volume} {62}},\ \bibinfo
  {pages} {043519} (\bibinfo {year} {2000})}\BibitemShut {NoStop}%
\bibitem [{\citenamefont {Chavanis}(2011)}]{chavanis1}%
  \BibitemOpen
  \bibfield  {author} {\bibinfo {author} {\bibfnamefont {P.-H.}\ \bibnamefont
  {Chavanis}},\ }\bibfield  {title} {\bibinfo {title} {{Mass-radius relation of
  Newtonian self-gravitating Bose-Einstein condensates with short-range
  interactions. I. Analytical results}},\ }\href
  {https://doi.org/10.1103/PhysRevD.84.043531} {\bibfield  {journal} {\bibinfo
  {journal} {Phys. Rev. D}\ }\textbf {\bibinfo {volume} {84}},\ \bibinfo
  {pages} {043531} (\bibinfo {year} {2011})}\BibitemShut {NoStop}%
\bibitem [{\citenamefont {Chavanis}\ and\ \citenamefont
  {Delfini}(2011)}]{chavanis2}%
  \BibitemOpen
  \bibfield  {author} {\bibinfo {author} {\bibfnamefont {P.-H.}\ \bibnamefont
  {Chavanis}}\ and\ \bibinfo {author} {\bibfnamefont {L.}~\bibnamefont
  {Delfini}},\ }\bibfield  {title} {\bibinfo {title} {{Mass-radius relation of
  Newtonian self-gravitating Bose-Einstein condensates with short-range
  interactions. II. Numerical results}},\ }\href
  {https://doi.org/10.1103/PhysRevD.84.043532} {\bibfield  {journal} {\bibinfo
  {journal} {Phys. Rev. D}\ }\textbf {\bibinfo {volume} {84}},\ \bibinfo
  {pages} {043532} (\bibinfo {year} {2011})}\BibitemShut {NoStop}%
\bibitem [{\citenamefont {Guth}\ \emph {et~al.}(2015)\citenamefont {Guth},
  \citenamefont {Hertzberg},\ and\ \citenamefont
  {Prescod-Weinstein}}]{GuthBEC}%
  \BibitemOpen
  \bibfield  {author} {\bibinfo {author} {\bibfnamefont {A.~H.}\ \bibnamefont
  {Guth}}, \bibinfo {author} {\bibfnamefont {M.~P.}\ \bibnamefont
  {Hertzberg}},\ and\ \bibinfo {author} {\bibfnamefont {C.}~\bibnamefont
  {Prescod-Weinstein}},\ }\bibfield  {title} {\bibinfo {title} {Do dark matter
  axions form a condensate with long-range correlation?},\ }\href
  {https://doi.org/10.1103/PhysRevD.92.103513} {\bibfield  {journal} {\bibinfo
  {journal} {Phys. Rev. D}\ }\textbf {\bibinfo {volume} {92}},\ \bibinfo
  {pages} {103513} (\bibinfo {year} {2015})}\BibitemShut {NoStop}%
\bibitem [{\citenamefont {Kaup}(1968)}]{Kaup1968}%
  \BibitemOpen
  \bibfield  {author} {\bibinfo {author} {\bibfnamefont {D.~J.}\ \bibnamefont
  {Kaup}},\ }\bibfield  {title} {\bibinfo {title} {{Klein-Gordon Geon}},\
  }\href {https://doi.org/10.1103/PhysRev.172.1331} {\bibfield  {journal}
  {\bibinfo  {journal} {Phys. Rev.}\ }\textbf {\bibinfo {volume} {172}},\
  \bibinfo {pages} {1331} (\bibinfo {year} {1968})}\BibitemShut {NoStop}%
\bibitem [{\citenamefont {Ruffini}\ and\ \citenamefont
  {Bonazzola}(1969)}]{Ruffini69}%
  \BibitemOpen
  \bibfield  {author} {\bibinfo {author} {\bibfnamefont {R.}~\bibnamefont
  {Ruffini}}\ and\ \bibinfo {author} {\bibfnamefont {S.}~\bibnamefont
  {Bonazzola}},\ }\bibfield  {title} {\bibinfo {title} {{Systems of
  self-gravitating particles in general relativity and the concept of an
  equation of state}},\ }\href {https://doi.org/10.1103/PhysRev.187.1767}
  {\bibfield  {journal} {\bibinfo  {journal} {Phys. Rev.}\ }\textbf {\bibinfo
  {volume} {187}},\ \bibinfo {pages} {1767} (\bibinfo {year}
  {1969})}\BibitemShut {NoStop}%
\bibitem [{\citenamefont {Breit}\ \emph {et~al.}(1984)\citenamefont {Breit},
  \citenamefont {Gupta},\ and\ \citenamefont {Zaks}}]{BREIT1984329}%
  \BibitemOpen
  \bibfield  {author} {\bibinfo {author} {\bibfnamefont {J.}~\bibnamefont
  {Breit}}, \bibinfo {author} {\bibfnamefont {S.}~\bibnamefont {Gupta}},\ and\
  \bibinfo {author} {\bibfnamefont {A.}~\bibnamefont {Zaks}},\ }\bibfield
  {title} {\bibinfo {title} {Cold {Bose} stars},\ }\href
  {https://doi.org/https://doi.org/10.1016/0370-2693(84)90764-0} {\bibfield
  {journal} {\bibinfo  {journal} {Physics Letters B}\ }\textbf {\bibinfo
  {volume} {140}},\ \bibinfo {pages} {329 } (\bibinfo {year}
  {1984})}\BibitemShut {NoStop}%
\bibitem [{\citenamefont {Urena-Lopez}\ \emph {et~al.}(2002)\citenamefont
  {Urena-Lopez}, \citenamefont {Matos},\ and\ \citenamefont
  {Becerril}}]{UrenaLopez2002}%
  \BibitemOpen
  \bibfield  {author} {\bibinfo {author} {\bibfnamefont {L.}~\bibnamefont
  {Urena-Lopez}}, \bibinfo {author} {\bibfnamefont {T.}~\bibnamefont {Matos}},\
  and\ \bibinfo {author} {\bibfnamefont {R.}~\bibnamefont {Becerril}},\
  }\bibfield  {title} {\bibinfo {title} {{Inside oscillatons}},\ }\href
  {https://doi.org/10.1088/0264-9381/19/23/320} {\bibfield  {journal} {\bibinfo
   {journal} {Class. Quant. Grav.}\ }\textbf {\bibinfo {volume} {19}},\
  \bibinfo {pages} {6259} (\bibinfo {year} {2002})}\BibitemShut {NoStop}%
\bibitem [{\citenamefont {Barranco}\ \emph {et~al.}(2013)\citenamefont
  {Barranco}, \citenamefont {Monteverde},\ and\ \citenamefont
  {Delepine}}]{Barranco2013}%
  \BibitemOpen
  \bibfield  {author} {\bibinfo {author} {\bibfnamefont {J.}~\bibnamefont
  {Barranco}}, \bibinfo {author} {\bibfnamefont {A.~C.}\ \bibnamefont
  {Monteverde}},\ and\ \bibinfo {author} {\bibfnamefont {D.}~\bibnamefont
  {Delepine}},\ }\bibfield  {title} {\bibinfo {title} {Can the dark matter halo
  be a collisionless ensemble of axion stars?},\ }\href
  {https://doi.org/10.1103/PhysRevD.87.103011} {\bibfield  {journal} {\bibinfo
  {journal} {Phys. Rev. D}\ }\textbf {\bibinfo {volume} {87}},\ \bibinfo
  {pages} {103011} (\bibinfo {year} {2013})}\BibitemShut {NoStop}%
\bibitem [{\citenamefont {Visinelli}\ \emph {et~al.}(2018)\citenamefont
  {Visinelli}, \citenamefont {Baum}, \citenamefont {Redondo}, \citenamefont
  {Freese},\ and\ \citenamefont {Wilczek}}]{baum2017}%
  \BibitemOpen
  \bibfield  {author} {\bibinfo {author} {\bibfnamefont {L.}~\bibnamefont
  {Visinelli}}, \bibinfo {author} {\bibfnamefont {S.}~\bibnamefont {Baum}},
  \bibinfo {author} {\bibfnamefont {J.}~\bibnamefont {Redondo}}, \bibinfo
  {author} {\bibfnamefont {K.}~\bibnamefont {Freese}},\ and\ \bibinfo {author}
  {\bibfnamefont {F.}~\bibnamefont {Wilczek}},\ }\bibfield  {title} {\bibinfo
  {title} {{Dilute and dense axion stars}},\ }\href
  {https://doi.org/10.1016/j.physletb.2017.12.010} {\bibfield  {journal}
  {\bibinfo  {journal} {Phys. Lett. B}\ }\textbf {\bibinfo {volume} {777}},\
  \bibinfo {pages} {64} (\bibinfo {year} {2018})},\ \Eprint
  {https://arxiv.org/abs/1710.08910} {arXiv:1710.08910 [astro-ph.CO]}
  \BibitemShut {NoStop}%
\bibitem [{\citenamefont {Barranco}\ and\ \citenamefont
  {Bernal}(2011)}]{Bernal2011}%
  \BibitemOpen
  \bibfield  {author} {\bibinfo {author} {\bibfnamefont {J.}~\bibnamefont
  {Barranco}}\ and\ \bibinfo {author} {\bibfnamefont {A.}~\bibnamefont
  {Bernal}},\ }\bibfield  {title} {\bibinfo {title} {Self-gravitating system
  made of axions},\ }\href {https://doi.org/10.1103/PhysRevD.83.043525}
  {\bibfield  {journal} {\bibinfo  {journal} {Phys. Rev. D}\ }\textbf {\bibinfo
  {volume} {83}},\ \bibinfo {pages} {043525} (\bibinfo {year}
  {2011})}\BibitemShut {NoStop}%
\bibitem [{\citenamefont {Kling}\ and\ \citenamefont
  {Rajaraman}(2017)}]{Kling1}%
  \BibitemOpen
  \bibfield  {author} {\bibinfo {author} {\bibfnamefont {F.}~\bibnamefont
  {Kling}}\ and\ \bibinfo {author} {\bibfnamefont {A.}~\bibnamefont
  {Rajaraman}},\ }\bibfield  {title} {\bibinfo {title} {Towards an analytic
  construction of the wavefunction of boson stars},\ }\href
  {https://doi.org/10.1103/PhysRevD.96.044039} {\bibfield  {journal} {\bibinfo
  {journal} {Phys. Rev. D}\ }\textbf {\bibinfo {volume} {96}},\ \bibinfo
  {pages} {044039} (\bibinfo {year} {2017})}\BibitemShut {NoStop}%
\bibitem [{\citenamefont {Kling}\ and\ \citenamefont
  {Rajaraman}(2018)}]{Kling2}%
  \BibitemOpen
  \bibfield  {author} {\bibinfo {author} {\bibfnamefont {F.}~\bibnamefont
  {Kling}}\ and\ \bibinfo {author} {\bibfnamefont {A.}~\bibnamefont
  {Rajaraman}},\ }\bibfield  {title} {\bibinfo {title} {Profiles of boson stars
  with self-interactions},\ }\href {https://doi.org/10.1103/PhysRevD.97.063012}
  {\bibfield  {journal} {\bibinfo  {journal} {Phys. Rev. D}\ }\textbf {\bibinfo
  {volume} {97}},\ \bibinfo {pages} {063012} (\bibinfo {year}
  {2018})}\BibitemShut {NoStop}%
\bibitem [{\citenamefont {Braaten}\ \emph {et~al.}(2016)\citenamefont
  {Braaten}, \citenamefont {Mohapatra},\ and\ \citenamefont
  {Zhang}}]{braatenDense2016}%
  \BibitemOpen
  \bibfield  {author} {\bibinfo {author} {\bibfnamefont {E.}~\bibnamefont
  {Braaten}}, \bibinfo {author} {\bibfnamefont {A.}~\bibnamefont {Mohapatra}},\
  and\ \bibinfo {author} {\bibfnamefont {H.}~\bibnamefont {Zhang}},\ }\bibfield
   {title} {\bibinfo {title} {Dense axion stars},\ }\href
  {https://doi.org/10.1103/PhysRevLett.117.121801} {\bibfield  {journal}
  {\bibinfo  {journal} {Phys. Rev. Lett.}\ }\textbf {\bibinfo {volume} {117}},\
  \bibinfo {pages} {121801} (\bibinfo {year} {2016})}\BibitemShut {NoStop}%
\bibitem [{\citenamefont {Chavanis}(2018)}]{chavanis3}%
  \BibitemOpen
  \bibfield  {author} {\bibinfo {author} {\bibfnamefont {P.-H.}\ \bibnamefont
  {Chavanis}},\ }\bibfield  {title} {\bibinfo {title} {Phase transitions
  between dilute and dense axion stars},\ }\href
  {https://doi.org/10.1103/PhysRevD.98.023009} {\bibfield  {journal} {\bibinfo
  {journal} {Phys. Rev. D}\ }\textbf {\bibinfo {volume} {98}},\ \bibinfo
  {pages} {023009} (\bibinfo {year} {2018})}\BibitemShut {NoStop}%
\bibitem [{\citenamefont {Amin}\ and\ \citenamefont
  {Mocz}(2019)}]{MustafaClustering}%
  \BibitemOpen
  \bibfield  {author} {\bibinfo {author} {\bibfnamefont {M.~A.}\ \bibnamefont
  {Amin}}\ and\ \bibinfo {author} {\bibfnamefont {P.}~\bibnamefont {Mocz}},\
  }\bibfield  {title} {\bibinfo {title} {Formation, gravitational clustering,
  and interactions of nonrelativistic solitons in an expanding universe},\
  }\href {https://doi.org/10.1103/PhysRevD.100.063507} {\bibfield  {journal}
  {\bibinfo  {journal} {Phys. Rev. D}\ }\textbf {\bibinfo {volume} {100}},\
  \bibinfo {pages} {063507} (\bibinfo {year} {2019})}\BibitemShut {NoStop}%
\bibitem [{\citenamefont {Oll{\'{e}}}\ \emph {et~al.}(2020)\citenamefont
  {Oll{\'{e}}}, \citenamefont {Pujol{\`{a}}s},\ and\ \citenamefont
  {Rompineve}}]{OscillonDarkMatter}%
  \BibitemOpen
  \bibfield  {author} {\bibinfo {author} {\bibfnamefont {J.}~\bibnamefont
  {Oll{\'{e}}}}, \bibinfo {author} {\bibfnamefont {O.}~\bibnamefont
  {Pujol{\`{a}}s}},\ and\ \bibinfo {author} {\bibfnamefont {F.}~\bibnamefont
  {Rompineve}},\ }\bibfield  {title} {\bibinfo {title} {Oscillons and dark
  matter},\ }\href {https://doi.org/10.1088/1475-7516/2020/02/006} {\bibfield
  {journal} {\bibinfo  {journal} {Journal of Cosmology and Astroparticle
  Physics}\ }\textbf {\bibinfo {volume} {2020}}\bibinfo  {number} { (02)},\
  \bibinfo {pages} {006}}\BibitemShut {NoStop}%
\bibitem [{\citenamefont {Eby}\ \emph {et~al.}(2016)\citenamefont {Eby},
  \citenamefont {Suranyi},\ and\ \citenamefont {Wijewardhana}}]{EbyLifetime}%
  \BibitemOpen
\bibfield  {number} {  }\bibfield  {author} {\bibinfo {author} {\bibfnamefont
  {J.}~\bibnamefont {Eby}}, \bibinfo {author} {\bibfnamefont {P.}~\bibnamefont
  {Suranyi}},\ and\ \bibinfo {author} {\bibfnamefont {L.~C.~R.}\ \bibnamefont
  {Wijewardhana}},\ }\bibfield  {title} {\bibinfo {title} {The lifetime of
  axion stars},\ }\href {https://doi.org/10.1142/S0217732316500905} {\bibfield
  {journal} {\bibinfo  {journal} {Modern Physics Letters A}\ }\textbf {\bibinfo
  {volume} {31}},\ \bibinfo {pages} {1650090} (\bibinfo {year} {2016})},\
  \Eprint {https://arxiv.org/abs/https://doi.org/10.1142/S0217732316500905}
  {https://doi.org/10.1142/S0217732316500905} \BibitemShut {NoStop}%
\bibitem [{\citenamefont {Eby}\ \emph {et~al.}(2018)\citenamefont {Eby},
  \citenamefont {Ma}, \citenamefont {Suranyi},\ and\ \citenamefont
  {Wijewardhana}}]{EbyDecayCondensate}%
  \BibitemOpen
  \bibfield  {author} {\bibinfo {author} {\bibfnamefont {J.}~\bibnamefont
  {Eby}}, \bibinfo {author} {\bibfnamefont {M.}~\bibnamefont {Ma}}, \bibinfo
  {author} {\bibfnamefont {P.}~\bibnamefont {Suranyi}},\ and\ \bibinfo {author}
  {\bibfnamefont {L.~C.~R.}\ \bibnamefont {Wijewardhana}},\ }\bibfield  {title}
  {\bibinfo {title} {Decay of ultralight axion condensates},\ }\href
  {https://doi.org/10.1007/JHEP01(2018)066} {\bibfield  {journal} {\bibinfo
  {journal} {Journal of High Energy Physics}\ }\textbf {\bibinfo {volume}
  {2018}},\ \bibinfo {pages} {66} (\bibinfo {year} {2018})}\BibitemShut
  {NoStop}%
\bibitem [{\citenamefont {Geng}(2020)}]{GengFlatten}%
  \BibitemOpen
  \bibfield  {author} {\bibinfo {author} {\bibfnamefont {H.}~\bibnamefont
  {Geng}},\ }\bibfield  {title} {\bibinfo {title} {{A Potential Mechanism for
  Inflation from Swampland Conjectures}},\ }\href
  {https://doi.org/10.1016/j.physletb.2020.135430} {\bibfield  {journal}
  {\bibinfo  {journal} {Phys. Lett. B}\ }\textbf {\bibinfo {volume} {805}},\
  \bibinfo {pages} {135430} (\bibinfo {year} {2020})},\ \Eprint
  {https://arxiv.org/abs/1910.14047} {arXiv:1910.14047 [hep-th]} \BibitemShut
  {NoStop}%
\bibitem [{\citenamefont {Dubovsky}\ \emph {et~al.}(2012)\citenamefont
  {Dubovsky}, \citenamefont {Lawrence},\ and\ \citenamefont
  {Roberts}}]{Dubovsky_2012}%
  \BibitemOpen
  \bibfield  {author} {\bibinfo {author} {\bibfnamefont {S.}~\bibnamefont
  {Dubovsky}}, \bibinfo {author} {\bibfnamefont {A.}~\bibnamefont {Lawrence}},\
  and\ \bibinfo {author} {\bibfnamefont {M.~M.}\ \bibnamefont {Roberts}},\
  }\bibfield  {title} {\bibinfo {title} {Axion monodromy in a model of
  holographic gluodynamics},\ }\bibfield  {journal} {\bibinfo  {journal}
  {Journal of High Energy Physics}\ }\textbf {\bibinfo {volume} {2012}},\ \href
  {https://doi.org/10.1007/jhep02(2012)053} {10.1007/jhep02(2012)053} (\bibinfo
  {year} {2012})\BibitemShut {NoStop}%
\bibitem [{\citenamefont {Silverstein}\ and\ \citenamefont
  {Westphal}(2008)}]{Eva2008}%
  \BibitemOpen
  \bibfield  {author} {\bibinfo {author} {\bibfnamefont {E.}~\bibnamefont
  {Silverstein}}\ and\ \bibinfo {author} {\bibfnamefont {A.}~\bibnamefont
  {Westphal}},\ }\bibfield  {title} {\bibinfo {title} {Monodromy in the cmb:
  Gravity waves and string inflation},\ }\href
  {https://doi.org/10.1103/PhysRevD.78.106003} {\bibfield  {journal} {\bibinfo
  {journal} {Phys. Rev. D}\ }\textbf {\bibinfo {volume} {78}},\ \bibinfo
  {pages} {106003} (\bibinfo {year} {2008})}\BibitemShut {NoStop}%
\bibitem [{\citenamefont {McAllister}\ \emph {et~al.}(2010)\citenamefont
  {McAllister}, \citenamefont {Silverstein},\ and\ \citenamefont
  {Westphal}}]{EvaMcAllister2008}%
  \BibitemOpen
  \bibfield  {author} {\bibinfo {author} {\bibfnamefont {L.}~\bibnamefont
  {McAllister}}, \bibinfo {author} {\bibfnamefont {E.}~\bibnamefont
  {Silverstein}},\ and\ \bibinfo {author} {\bibfnamefont {A.}~\bibnamefont
  {Westphal}},\ }\bibfield  {title} {\bibinfo {title} {Gravity waves and linear
  inflation from axion monodromy},\ }\href
  {https://doi.org/10.1103/PhysRevD.82.046003} {\bibfield  {journal} {\bibinfo
  {journal} {Phys. Rev. D}\ }\textbf {\bibinfo {volume} {82}},\ \bibinfo
  {pages} {046003} (\bibinfo {year} {2010})}\BibitemShut {NoStop}%
\bibitem [{\citenamefont {Amin}\ \emph {et~al.}(2012)\citenamefont {Amin},
  \citenamefont {Easther}, \citenamefont {Finkel}, \citenamefont {Flauger},\
  and\ \citenamefont {Hertzberg}}]{MustafaOscillon}%
  \BibitemOpen
  \bibfield  {author} {\bibinfo {author} {\bibfnamefont {M.~A.}\ \bibnamefont
  {Amin}}, \bibinfo {author} {\bibfnamefont {R.}~\bibnamefont {Easther}},
  \bibinfo {author} {\bibfnamefont {H.}~\bibnamefont {Finkel}}, \bibinfo
  {author} {\bibfnamefont {R.}~\bibnamefont {Flauger}},\ and\ \bibinfo {author}
  {\bibfnamefont {M.~P.}\ \bibnamefont {Hertzberg}},\ }\bibfield  {title}
  {\bibinfo {title} {Oscillons after inflation},\ }\href
  {https://doi.org/10.1103/PhysRevLett.108.241302} {\bibfield  {journal}
  {\bibinfo  {journal} {Phys. Rev. Lett.}\ }\textbf {\bibinfo {volume} {108}},\
  \bibinfo {pages} {241302} (\bibinfo {year} {2012})}\BibitemShut {NoStop}%
\bibitem [{\citenamefont {Lozanov}\ and\ \citenamefont
  {Amin}(2018)}]{LozanovOscillon}%
  \BibitemOpen
  \bibfield  {author} {\bibinfo {author} {\bibfnamefont {K.~D.}\ \bibnamefont
  {Lozanov}}\ and\ \bibinfo {author} {\bibfnamefont {M.~A.}\ \bibnamefont
  {Amin}},\ }\bibfield  {title} {\bibinfo {title} {Self-resonance after
  inflation: Oscillons, transients, and radiation domination},\ }\href
  {https://doi.org/10.1103/PhysRevD.97.023533} {\bibfield  {journal} {\bibinfo
  {journal} {Phys. Rev. D}\ }\textbf {\bibinfo {volume} {97}},\ \bibinfo
  {pages} {023533} (\bibinfo {year} {2018})}\BibitemShut {NoStop}%
\bibitem [{\citenamefont {Kallosh}\ and\ \citenamefont
  {Linde}(2013)}]{LindeKallosh}%
  \BibitemOpen
  \bibfield  {author} {\bibinfo {author} {\bibfnamefont {R.}~\bibnamefont
  {Kallosh}}\ and\ \bibinfo {author} {\bibfnamefont {A.}~\bibnamefont
  {Linde}},\ }\bibfield  {title} {\bibinfo {title} {{Universality Class in
  Conformal Inflation}},\ }\href
  {https://doi.org/10.1088/1475-7516/2013/07/002} {\bibfield  {journal}
  {\bibinfo  {journal} {JCAP}\ }\textbf {\bibinfo {volume} {07}},\ \bibinfo
  {pages} {002}},\ \Eprint {https://arxiv.org/abs/1306.5220} {arXiv:1306.5220
  [hep-th]} \BibitemShut {NoStop}%
\bibitem [{\citenamefont {Zhang}\ \emph {et~al.}(2020)\citenamefont {Zhang},
  \citenamefont {Amin}, \citenamefont {Copeland}, \citenamefont {Saffin},\ and\
  \citenamefont {Lozanov}}]{MustafaOscillonDecay}%
  \BibitemOpen
  \bibfield  {author} {\bibinfo {author} {\bibfnamefont {H.-Y.}\ \bibnamefont
  {Zhang}}, \bibinfo {author} {\bibfnamefont {M.~A.}\ \bibnamefont {Amin}},
  \bibinfo {author} {\bibfnamefont {E.~J.}\ \bibnamefont {Copeland}}, \bibinfo
  {author} {\bibfnamefont {P.~M.}\ \bibnamefont {Saffin}},\ and\ \bibinfo
  {author} {\bibfnamefont {K.~D.}\ \bibnamefont {Lozanov}},\ }\bibfield
  {title} {\bibinfo {title} {{Classical Decay Rates of Oscillons}},\ }\href
  {https://doi.org/10.1088/1475-7516/2020/07/055} {\bibfield  {journal}
  {\bibinfo  {journal} {JCAP}\ }\textbf {\bibinfo {volume} {07}},\ \bibinfo
  {pages} {055}},\ \Eprint {https://arxiv.org/abs/2004.01202} {arXiv:2004.01202
  [hep-th]} \BibitemShut {NoStop}%
\bibitem [{\citenamefont {Kawasaki}\ \emph {et~al.}(2021)\citenamefont
  {Kawasaki}, \citenamefont {Nakano}, \citenamefont {Nakatsuka},\ and\
  \citenamefont {Sonomoto}}]{Kawasaki2020}%
  \BibitemOpen
  \bibfield  {author} {\bibinfo {author} {\bibfnamefont {M.}~\bibnamefont
  {Kawasaki}}, \bibinfo {author} {\bibfnamefont {W.}~\bibnamefont {Nakano}},
  \bibinfo {author} {\bibfnamefont {H.}~\bibnamefont {Nakatsuka}},\ and\
  \bibinfo {author} {\bibfnamefont {E.}~\bibnamefont {Sonomoto}},\ }\bibfield
  {title} {\bibinfo {title} {{Oscillons of Axion-Like Particle: Mass
  distribution and power spectrum}},\ }\href
  {https://doi.org/10.1088/1475-7516/2021/01/061} {\bibfield  {journal}
  {\bibinfo  {journal} {JCAP}\ }\textbf {\bibinfo {volume} {01}},\ \bibinfo
  {pages} {061}},\ \Eprint {https://arxiv.org/abs/2010.09311} {arXiv:2010.09311
  [astro-ph.CO]} \BibitemShut {NoStop}%
\bibitem [{\citenamefont {Cyncynates}\ and\ \citenamefont
  {Giurgica-Tiron}(2021)}]{Cyncynates2021}%
  \BibitemOpen
  \bibfield  {author} {\bibinfo {author} {\bibfnamefont {D.}~\bibnamefont
  {Cyncynates}}\ and\ \bibinfo {author} {\bibfnamefont {T.}~\bibnamefont
  {Giurgica-Tiron}},\ }\bibfield  {title} {\bibinfo {title} {{The Structure of
  the Oscillon: The Dynamics of Attractive Self-Interaction}},\ }\href@noop {}
  {\  (\bibinfo {year} {2021})},\ \Eprint {https://arxiv.org/abs/2104.02069}
  {arXiv:2104.02069 [hep-ph]} \BibitemShut {NoStop}%
\bibitem [{\citenamefont {Tilburg}\ \emph {et~al.}(2018)\citenamefont
  {Tilburg}, \citenamefont {Taki},\ and\ \citenamefont
  {Weiner}}]{kenAstrometry}%
  \BibitemOpen
  \bibfield  {author} {\bibinfo {author} {\bibfnamefont {K.~V.}\ \bibnamefont
  {Tilburg}}, \bibinfo {author} {\bibfnamefont {A.-M.}\ \bibnamefont {Taki}},\
  and\ \bibinfo {author} {\bibfnamefont {N.}~\bibnamefont {Weiner}},\
  }\bibfield  {title} {\bibinfo {title} {Halometry from astrometry},\ }\href
  {https://doi.org/10.1088/1475-7516/2018/07/041} {\bibfield  {journal}
  {\bibinfo  {journal} {Journal of Cosmology and Astroparticle Physics}\
  }\textbf {\bibinfo {volume} {2018}}\bibinfo  {number} { (07)},\ \bibinfo
  {pages} {041}}\BibitemShut {NoStop}%
\bibitem [{\citenamefont {Prabhu}()}]{prabhu2}%
  \BibitemOpen
\bibfield  {number} {  }\bibfield  {author} {\bibinfo {author} {\bibfnamefont
  {A.}~\bibnamefont {Prabhu}},\ }\href@noop {} {\bibinfo {title} {Optical
  lensing by axion stars {II}: Lensing signatures of dense axion
  configurations}},\ \bibinfo {note} {in preparation}\BibitemShut {NoStop}%
\bibitem [{\citenamefont {Prusti}\ \emph {et~al.}(2016)\citenamefont {Prusti}
  \emph {et~al.}}]{Gaia2016}%
  \BibitemOpen
  \bibfield  {author} {\bibinfo {author} {\bibfnamefont {T.}~\bibnamefont
  {Prusti}} \emph {et~al.} (\bibinfo {collaboration} {{Gaia}}),\ }\bibfield
  {title} {\bibinfo {title} {{The Gaia Mission}},\ }\href
  {https://doi.org/10.1051/0004-6361/201629272} {\bibfield  {journal} {\bibinfo
   {journal} {Astron. Astrophys.}\ }\textbf {\bibinfo {volume} {595}},\
  \bibinfo {pages} {A1} (\bibinfo {year} {2016})},\ \Eprint
  {https://arxiv.org/abs/1609.04153} {arXiv:1609.04153 [astro-ph.IM]}
  \BibitemShut {NoStop}%
\bibitem [{\citenamefont {Fomalont}\ and\ \citenamefont
  {Reid}(2004)}]{fomalont}%
  \BibitemOpen
  \bibfield  {author} {\bibinfo {author} {\bibfnamefont {E.}~\bibnamefont
  {Fomalont}}\ and\ \bibinfo {author} {\bibfnamefont {M.}~\bibnamefont
  {Reid}},\ }\bibfield  {title} {\bibinfo {title} {Microarcsecond astrometry
  using the {SKA}},\ }\href
  {https://doi.org/https://doi.org/10.1016/j.newar.2004.09.037} {\bibfield
  {journal} {\bibinfo  {journal} {New Astronomy Reviews}\ }\textbf {\bibinfo
  {volume} {48}},\ \bibinfo {pages} {1473 } (\bibinfo {year} {2004})},\
  \bibinfo {note} {science with the Square Kilometre Array}\BibitemShut
  {NoStop}%
\bibitem [{\citenamefont {{Charlot}}(2012)}]{VLBIGaia2012}%
  \BibitemOpen
  \bibfield  {author} {\bibinfo {author} {\bibfnamefont {P.}~\bibnamefont
  {{Charlot}}},\ }\bibfield  {title} {\bibinfo {title} {{Precision Astrometry:
  From VLBI to Gaia and SKA}},\ }\href {https://doi.org/K13-85497} {\bibfield
  {journal} {\bibinfo  {journal} {Astrophysics and Space Science Proceedings}\
  }\textbf {\bibinfo {volume} {25}},\ \bibinfo {pages} {85} (\bibinfo {year}
  {2012})}\BibitemShut {NoStop}%
\bibitem [{\citenamefont {{Navarro}}\ \emph {et~al.}(1997)\citenamefont
  {{Navarro}}, \citenamefont {{Frenk}},\ and\ \citenamefont {{White}}}]{NFW}%
  \BibitemOpen
  \bibfield  {author} {\bibinfo {author} {\bibfnamefont {J.~F.}\ \bibnamefont
  {{Navarro}}}, \bibinfo {author} {\bibfnamefont {C.~S.}\ \bibnamefont
  {{Frenk}}},\ and\ \bibinfo {author} {\bibfnamefont {S.~D.~M.}\ \bibnamefont
  {{White}}},\ }\bibfield  {title} {\bibinfo {title} {{A Universal Density
  Profile from Hierarchical Clustering}},\ }\href
  {https://doi.org/10.1086/304888} {\bibfield  {journal} {\bibinfo  {journal}
  {\apj}\ }\textbf {\bibinfo {volume} {490}},\ \bibinfo {pages} {493} (\bibinfo
  {year} {1997})},\ \Eprint {https://arxiv.org/abs/astro-ph/9611107}
  {arXiv:astro-ph/9611107 [astro-ph]} \BibitemShut {NoStop}%
\bibitem [{\citenamefont {{Moore}}(1994)}]{CuspCore94}%
  \BibitemOpen
  \bibfield  {author} {\bibinfo {author} {\bibfnamefont {B.}~\bibnamefont
  {{Moore}}},\ }\bibfield  {title} {\bibinfo {title} {{Evidence against
  dissipation-less dark matter from observations of galaxy haloes}},\ }\href
  {https://doi.org/10.1038/370629a0} {\bibfield  {journal} {\bibinfo  {journal}
  {\nat}\ }\textbf {\bibinfo {volume} {370}},\ \bibinfo {pages} {629} (\bibinfo
  {year} {1994})}\BibitemShut {NoStop}%
\bibitem [{\citenamefont {Oh}\ \emph {et~al.}(2015)\citenamefont {Oh} \emph
  {et~al.}}]{CuspCore2015}%
  \BibitemOpen
  \bibfield  {author} {\bibinfo {author} {\bibfnamefont {S.-H.}\ \bibnamefont
  {Oh}} \emph {et~al.},\ }\bibfield  {title} {\bibinfo {title}
  {{High-resolution mass models of dwarf galaxies from LITTLE THINGS}},\ }\href
  {https://doi.org/10.1088/0004-6256/149/6/180} {\bibfield  {journal} {\bibinfo
   {journal} {Astron. J.}\ }\textbf {\bibinfo {volume} {149}},\ \bibinfo
  {pages} {180} (\bibinfo {year} {2015})},\ \Eprint
  {https://arxiv.org/abs/1502.01281} {arXiv:1502.01281 [astro-ph.GA]}
  \BibitemShut {NoStop}%
\bibitem [{\citenamefont {Kim}(1979)}]{KimOG}%
  \BibitemOpen
  \bibfield  {author} {\bibinfo {author} {\bibfnamefont {J.~E.}\ \bibnamefont
  {Kim}},\ }\bibfield  {title} {\bibinfo {title} {Weak-interaction singlet and
  strong $\mathrm{CP}$ invariance},\ }\href
  {https://doi.org/10.1103/PhysRevLett.43.103} {\bibfield  {journal} {\bibinfo
  {journal} {Phys. Rev. Lett.}\ }\textbf {\bibinfo {volume} {43}},\ \bibinfo
  {pages} {103} (\bibinfo {year} {1979})}\BibitemShut {NoStop}%
\bibitem [{\citenamefont {Shifman}\ \emph {et~al.}(1980)\citenamefont
  {Shifman}, \citenamefont {Vainshtein},\ and\ \citenamefont
  {Zakharov}}]{SVZOG}%
  \BibitemOpen
  \bibfield  {author} {\bibinfo {author} {\bibfnamefont {M.}~\bibnamefont
  {Shifman}}, \bibinfo {author} {\bibfnamefont {A.}~\bibnamefont
  {Vainshtein}},\ and\ \bibinfo {author} {\bibfnamefont {V.}~\bibnamefont
  {Zakharov}},\ }\bibfield  {title} {\bibinfo {title} {Can confinement ensure
  natural {CP} invariance of strong interactions?},\ }\href
  {https://doi.org/https://doi.org/10.1016/0550-3213(80)90209-6} {\bibfield
  {journal} {\bibinfo  {journal} {Nuclear Physics B}\ }\textbf {\bibinfo
  {volume} {166}},\ \bibinfo {pages} {493 } (\bibinfo {year}
  {1980})}\BibitemShut {NoStop}%
\bibitem [{\citenamefont {Dine}\ \emph {et~al.}(1981)\citenamefont {Dine},
  \citenamefont {Fischler},\ and\ \citenamefont {Srednicki}}]{DFSOG}%
  \BibitemOpen
  \bibfield  {author} {\bibinfo {author} {\bibfnamefont {M.}~\bibnamefont
  {Dine}}, \bibinfo {author} {\bibfnamefont {W.}~\bibnamefont {Fischler}},\
  and\ \bibinfo {author} {\bibfnamefont {M.}~\bibnamefont {Srednicki}},\
  }\bibfield  {title} {\bibinfo {title} {A simple solution to the strong {CP}
  problem with a harmless axion},\ }\href
  {https://doi.org/https://doi.org/10.1016/0370-2693(81)90590-6} {\bibfield
  {journal} {\bibinfo  {journal} {Physics Letters B}\ }\textbf {\bibinfo
  {volume} {104}},\ \bibinfo {pages} {199 } (\bibinfo {year}
  {1981})}\BibitemShut {NoStop}%
\bibitem [{\citenamefont {Zhitnitsky}(1980)}]{ZOG}%
  \BibitemOpen
  \bibfield  {author} {\bibinfo {author} {\bibfnamefont {A.}~\bibnamefont
  {Zhitnitsky}},\ }\bibfield  {title} {\bibinfo {title} {{On Possible
  Suppression of the Axion Hadron Interactions. (In Russian)}},\ }\href@noop {}
  {\bibfield  {journal} {\bibinfo  {journal} {Sov. J. Nucl. Phys.}\ }\textbf
  {\bibinfo {volume} {31}},\ \bibinfo {pages} {260} (\bibinfo {year}
  {1980})}\BibitemShut {NoStop}%
\bibitem [{\citenamefont {Di~Luzio}\ \emph
  {et~al.}(2017{\natexlab{a}})\citenamefont {Di~Luzio}, \citenamefont
  {Mescia},\ and\ \citenamefont {Nardi}}]{DiLuzio2017}%
  \BibitemOpen
  \bibfield  {author} {\bibinfo {author} {\bibfnamefont {L.}~\bibnamefont
  {Di~Luzio}}, \bibinfo {author} {\bibfnamefont {F.}~\bibnamefont {Mescia}},\
  and\ \bibinfo {author} {\bibfnamefont {E.}~\bibnamefont {Nardi}},\ }\bibfield
   {title} {\bibinfo {title} {Redefining the axion window},\ }\bibfield
  {journal} {\bibinfo  {journal} {Physical Review Letters}\ }\textbf {\bibinfo
  {volume} {118}},\ \href {https://doi.org/10.1103/physrevlett.118.031801}
  {10.1103/physrevlett.118.031801} (\bibinfo {year}
  {2017}{\natexlab{a}})\BibitemShut {NoStop}%
\bibitem [{\citenamefont {Di~Luzio}\ \emph
  {et~al.}(2017{\natexlab{b}})\citenamefont {Di~Luzio}, \citenamefont
  {Mescia},\ and\ \citenamefont {Nardi}}]{DiLuzio20172}%
  \BibitemOpen
  \bibfield  {author} {\bibinfo {author} {\bibfnamefont {L.}~\bibnamefont
  {Di~Luzio}}, \bibinfo {author} {\bibfnamefont {F.}~\bibnamefont {Mescia}},\
  and\ \bibinfo {author} {\bibfnamefont {E.}~\bibnamefont {Nardi}},\ }\bibfield
   {title} {\bibinfo {title} {Window for preferred axion models},\ }\bibfield
  {journal} {\bibinfo  {journal} {Physical Review D}\ }\textbf {\bibinfo
  {volume} {96}},\ \href {https://doi.org/10.1103/physrevd.96.075003}
  {10.1103/physrevd.96.075003} (\bibinfo {year}
  {2017}{\natexlab{b}})\BibitemShut {NoStop}%
\bibitem [{\citenamefont {Agrawal}\ \emph {et~al.}(2018)\citenamefont
  {Agrawal}, \citenamefont {Fan}, \citenamefont {Reece},\ and\ \citenamefont
  {Wang}}]{Prateek2018}%
  \BibitemOpen
  \bibfield  {author} {\bibinfo {author} {\bibfnamefont {P.}~\bibnamefont
  {Agrawal}}, \bibinfo {author} {\bibfnamefont {J.}~\bibnamefont {Fan}},
  \bibinfo {author} {\bibfnamefont {M.}~\bibnamefont {Reece}},\ and\ \bibinfo
  {author} {\bibfnamefont {L.-T.}\ \bibnamefont {Wang}},\ }\bibfield  {title}
  {\bibinfo {title} {Experimental targets for photon couplings of the qcd
  axion},\ }\bibfield  {journal} {\bibinfo  {journal} {Journal of High Energy
  Physics}\ }\textbf {\bibinfo {volume} {2018}},\ \href
  {https://doi.org/10.1007/jhep02(2018)006} {10.1007/jhep02(2018)006} (\bibinfo
  {year} {2018})\BibitemShut {NoStop}%
\bibitem [{\citenamefont {Press}\ and\ \citenamefont
  {Schechter}(1974)}]{PressSchechter}%
  \BibitemOpen
  \bibfield  {author} {\bibinfo {author} {\bibfnamefont {W.~H.}\ \bibnamefont
  {Press}}\ and\ \bibinfo {author} {\bibfnamefont {P.}~\bibnamefont
  {Schechter}},\ }\bibfield  {title} {\bibinfo {title} {{Formation of galaxies
  and clusters of galaxies by self-similar gravitational condensation}},\
  }\href {https://doi.org/10.1086/152650} {\bibfield  {journal} {\bibinfo
  {journal} {Astrophys. J.}\ }\textbf {\bibinfo {volume} {187}},\ \bibinfo
  {pages} {425} (\bibinfo {year} {1974})}\BibitemShut {NoStop}%
\bibitem [{\citenamefont {Hertzberg}\ and\ \citenamefont
  {Schiappacasse}(2018)}]{Hertzberg_2018}%
  \BibitemOpen
  \bibfield  {author} {\bibinfo {author} {\bibfnamefont {M.~P.}\ \bibnamefont
  {Hertzberg}}\ and\ \bibinfo {author} {\bibfnamefont {E.~D.}\ \bibnamefont
  {Schiappacasse}},\ }\bibfield  {title} {\bibinfo {title} {Dark matter axion
  clump resonance of photons},\ }\href
  {https://doi.org/10.1088/1475-7516/2018/11/004} {\bibfield  {journal}
  {\bibinfo  {journal} {Journal of Cosmology and Astroparticle Physics}\
  }\textbf {\bibinfo {volume} {2018}}\bibinfo  {number} { (11)},\ \bibinfo
  {pages} {004}}\BibitemShut {NoStop}%
\bibitem [{\citenamefont {Levkov}\ \emph {et~al.}(2020)\citenamefont {Levkov},
  \citenamefont {Panin},\ and\ \citenamefont
  {Tkachev}}]{levkov2020radioemission}%
  \BibitemOpen
\bibfield  {number} {  }\bibfield  {author} {\bibinfo {author} {\bibfnamefont
  {D.~G.}\ \bibnamefont {Levkov}}, \bibinfo {author} {\bibfnamefont {A.~G.}\
  \bibnamefont {Panin}},\ and\ \bibinfo {author} {\bibfnamefont {I.~I.}\
  \bibnamefont {Tkachev}},\ }\href@noop {} {\bibinfo {title} {Radio-emission of
  axion stars}} (\bibinfo {year} {2020}),\ \Eprint
  {https://arxiv.org/abs/2004.05179} {arXiv:2004.05179 [astro-ph.CO]}
  \BibitemShut {NoStop}%
\bibitem [{\citenamefont {Hertzberg}\ \emph {et~al.}(2020)\citenamefont
  {Hertzberg}, \citenamefont {Li},\ and\ \citenamefont
  {Schiappacasse}}]{Hertzberg_2020}%
  \BibitemOpen
  \bibfield  {author} {\bibinfo {author} {\bibfnamefont {M.~P.}\ \bibnamefont
  {Hertzberg}}, \bibinfo {author} {\bibfnamefont {Y.}~\bibnamefont {Li}},\ and\
  \bibinfo {author} {\bibfnamefont {E.~D.}\ \bibnamefont {Schiappacasse}},\
  }\bibfield  {title} {\bibinfo {title} {{Merger of Dark Matter Axion Clumps
  and Resonant Photon Emission}},\ }\href
  {https://doi.org/10.1088/1475-7516/2020/07/067} {\bibfield  {journal}
  {\bibinfo  {journal} {JCAP}\ }\textbf {\bibinfo {volume} {07}},\ \bibinfo
  {pages} {067}},\ \Eprint {https://arxiv.org/abs/2005.02405} {arXiv:2005.02405
  [hep-ph]} \BibitemShut {NoStop}%
\bibitem [{\citenamefont {Amin}\ and\ \citenamefont {Mou}(2021)}]{Amin2020vja}%
  \BibitemOpen
  \bibfield  {author} {\bibinfo {author} {\bibfnamefont {M.~A.}\ \bibnamefont
  {Amin}}\ and\ \bibinfo {author} {\bibfnamefont {Z.-G.}\ \bibnamefont {Mou}},\
  }\bibfield  {title} {\bibinfo {title} {{Electromagnetic Bursts from Mergers
  of Oscillons in Axion-like Fields}},\ }\href
  {https://doi.org/10.1088/1475-7516/2021/02/024} {\bibfield  {journal}
  {\bibinfo  {journal} {JCAP}\ }\textbf {\bibinfo {volume} {02}},\ \bibinfo
  {pages} {024}},\ \Eprint {https://arxiv.org/abs/2009.11337} {arXiv:2009.11337
  [astro-ph.CO]} \BibitemShut {NoStop}%
\bibitem [{\citenamefont {Amin}\ \emph {et~al.}(2021)\citenamefont {Amin},
  \citenamefont {Long}, \citenamefont {Mou},\ and\ \citenamefont
  {Saffin}}]{Amin2021tnq}%
  \BibitemOpen
  \bibfield  {author} {\bibinfo {author} {\bibfnamefont {M.~A.}\ \bibnamefont
  {Amin}}, \bibinfo {author} {\bibfnamefont {A.~J.}\ \bibnamefont {Long}},
  \bibinfo {author} {\bibfnamefont {Z.-G.}\ \bibnamefont {Mou}},\ and\ \bibinfo
  {author} {\bibfnamefont {P.~M.}\ \bibnamefont {Saffin}},\ }\bibfield  {title}
  {\bibinfo {title} {{Dipole Radiation and Beyond from Axion Stars in
  Electromagnetic Fields}},\ }\href@noop {} {\  (\bibinfo {year} {2021})},\
  \Eprint {https://arxiv.org/abs/2103.12082} {arXiv:2103.12082 [hep-ph]}
  \BibitemShut {NoStop}%
\bibitem [{\citenamefont {Carr}\ \emph
  {et~al.}(2010{\natexlab{b}})\citenamefont {Carr}, \citenamefont {Kohri},
  \citenamefont {Sendouda},\ and\ \citenamefont {Yokoyama}}]{PBHBBN2010}%
  \BibitemOpen
  \bibfield  {author} {\bibinfo {author} {\bibfnamefont {B.~J.}\ \bibnamefont
  {Carr}}, \bibinfo {author} {\bibfnamefont {K.}~\bibnamefont {Kohri}},
  \bibinfo {author} {\bibfnamefont {Y.}~\bibnamefont {Sendouda}},\ and\
  \bibinfo {author} {\bibfnamefont {J.}~\bibnamefont {Yokoyama}},\ }\bibfield
  {title} {\bibinfo {title} {New cosmological constraints on primordial black
  holes},\ }\href {https://doi.org/10.1103/PhysRevD.81.104019} {\bibfield
  {journal} {\bibinfo  {journal} {Phys. Rev. D}\ }\textbf {\bibinfo {volume}
  {81}},\ \bibinfo {pages} {104019} (\bibinfo {year}
  {2010}{\natexlab{b}})}\BibitemShut {NoStop}%
\bibitem [{\citenamefont {Barnacka}\ \emph {et~al.}(2012)\citenamefont
  {Barnacka}, \citenamefont {Glicenstein},\ and\ \citenamefont
  {Moderski}}]{femtolensing2012}%
  \BibitemOpen
  \bibfield  {author} {\bibinfo {author} {\bibfnamefont {A.}~\bibnamefont
  {Barnacka}}, \bibinfo {author} {\bibfnamefont {J.-F.}\ \bibnamefont
  {Glicenstein}},\ and\ \bibinfo {author} {\bibfnamefont {R.}~\bibnamefont
  {Moderski}},\ }\bibfield  {title} {\bibinfo {title} {New constraints on
  primordial black holes abundance from femtolensing of gamma-ray bursts},\
  }\href {https://doi.org/10.1103/PhysRevD.86.043001} {\bibfield  {journal}
  {\bibinfo  {journal} {Phys. Rev. D}\ }\textbf {\bibinfo {volume} {86}},\
  \bibinfo {pages} {043001} (\bibinfo {year} {2012})}\BibitemShut {NoStop}%
\bibitem [{\citenamefont {Griest}\ \emph {et~al.}(2014)\citenamefont {Griest},
  \citenamefont {Cieplak},\ and\ \citenamefont {Lehner}}]{Kepler2014}%
  \BibitemOpen
  \bibfield  {author} {\bibinfo {author} {\bibfnamefont {K.}~\bibnamefont
  {Griest}}, \bibinfo {author} {\bibfnamefont {A.~M.}\ \bibnamefont
  {Cieplak}},\ and\ \bibinfo {author} {\bibfnamefont {M.~J.}\ \bibnamefont
  {Lehner}},\ }\bibfield  {title} {\bibinfo {title} {{Experimental Limits on
  Primordial Black Hole Dark Matter from the First 2 yr of Kepler Data}},\
  }\href {https://doi.org/10.1088/0004-637X/786/2/158} {\bibfield  {journal}
  {\bibinfo  {journal} {Astrophys. J.}\ }\textbf {\bibinfo {volume} {786}},\
  \bibinfo {pages} {158} (\bibinfo {year} {2014})}\BibitemShut {NoStop}%
\end{thebibliography}%


\providecommand{\noopsort}[1]{}\providecommand{\singleletter}[1]{#1}%
%

\end{document}